\documentclass[final]{IEEEtran}
\usepackage{times,graphicx,float}
\usepackage{amsmath}
\usepackage{amssymb}

 \usepackage{color}
 \usepackage[normalem]{ulem}
 
 \newcommand\reviseA[1]{{\color{black}#1}}
  \newcommand{\tabincell}[2]{\begin{tabular}{@{}#1@{}}#2\end{tabular}}

\newcommand{\cA}{{\mathcal A}}

\newcommand{\bL}{{\boldsymbol L}}
\newcommand{\bR}{{\boldsymbol R}}

\newcommand{\bH}{{\boldsymbol H}}
\newcommand{\ba}{{\boldsymbol a}}
\newcommand{\bQ}{{\boldsymbol Q}}
\newcommand{\bA}{{\boldsymbol A}}
\newcommand{\bPsi}{{\boldsymbol \Psi}}

\newcommand{\bSigma}{{\boldsymbol \Sigma}}
\newcommand{\bb}{{\boldsymbol b}}
\newcommand{\br}{{\boldsymbol r}}
\newcommand{\bB}{{\boldsymbol B}}
 
\newcommand{\bC}{{\boldsymbol C}}
\newcommand{\bD}{{\boldsymbol D}}
\newcommand{\bE}{{\boldsymbol E}}
\newcommand{\bZ}{{\boldsymbol Z}}
\newcommand{\bc}{{\boldsymbol c}}

\newcommand{\bx}{{\boldsymbol x}}
\newcommand{\bz}{{\boldsymbol z}}

\newcommand{\cT}{{\mathcal T}}

\newcommand{\bu}{{\boldsymbol u}}

\newcommand{\bI}{{\boldsymbol I}}
\newcommand{\bY}{{\boldsymbol Y}}
\newcommand{\bV}{{\boldsymbol V}}
\newcommand{\bX}{{\boldsymbol X}}
\newcommand{\bN}{{\boldsymbol N}}
\newcommand{\bW}{{\boldsymbol W}}

 \DeclareMathOperator{\argmin}{argmin}
  \DeclareMathOperator{\argmax}{argmax}
 \DeclareMathOperator{\trace}{Tr}
  
 \DeclareMathOperator{\toep}{\mathcal{T}}
 \DeclareMathOperator{\diag}{diag}
  \DeclareMathOperator{\rank}{rank}
  
\usepackage{cite}\usepackage{amsthm}\usepackage{array}\usepackage{comment}
\usepackage{dsfont}
\usepackage{mathrsfs}
\graphicspath{{figures/} }

\usepackage[T1]{fontenc}
\usepackage[latin9]{inputenc}

\begin{document}
\theoremstyle{plain}\newtheorem{lemma}{\textbf{Lemma}}\newtheorem{theorem}{\textbf{Theorem}}\newtheorem{corollary}{\textbf{Corollary}}\newtheorem{assumption}{\textbf{Assumption}}\newtheorem{example}{\textbf{Example}}\newtheorem{definition}{\textbf{Definition}}
\newtheorem{prop}{\textbf{Proposition}}
\theoremstyle{definition}

\theoremstyle{remark}\newtheorem{remark}{\textbf{Remark}}

\title{Off-the-Grid Line Spectrum Denoising and Estimation with Multiple Measurement Vectors}
\author{Yuanxin Li and Yuejie Chi$^{\star}$\thanks{The authors are with Department of Electrical and Computer Engineering, The Ohio State University, Columbus, OH 43210 USA (e-mails: \{li.3822, chi.97\}@osu.edu). This work is supported in part by NSF under grant CCF-1422966, by AFOSR under grant FA9550-15-1-0205, by ONR under grant N00014-15-1-2387, and by the Ralph E. Powe Junior Faculty Enhancement Award from the Oak Ridge Associated Universities. Corresponding e-mail: chi.97@osu.edu. Date: \today.}\thanks{Parts of the results in this paper were presented at the IEEE International Conference on Acoustics, Speech and Signal Processing, Florence, Italy, May 2014 \cite{chi2013joint} and the Statistical Signal Processing Workshop, Gold Coast, Australia, June 2014 \cite{li2014compressive}. } }

\maketitle

\begin{abstract}
\reviseA{Compressed Sensing suggests that the required number of samples for reconstructing a signal can be greatly reduced if it is sparse in a known discrete basis, yet many real-world signals are sparse in a continuous dictionary.} One example is the spectrally-sparse signal, which is composed of a small number of spectral atoms with arbitrary frequencies on the unit interval. \reviseA{In this paper we study the problem of line spectrum denoising and estimation with an ensemble of spectrally-sparse signals composed of the same set of continuous-valued frequencies from their partial and noisy observations.} Two approaches are developed based on atomic norm minimization and {structured covariance estimation}, both of which can be solved efficiently via semidefinite programming. The first approach aims to estimate and denoise the set of signals from their partial and noisy observations via atomic norm minimization, and recover the frequencies via examining the dual polynomial of the convex program. We characterize the optimality condition of the proposed algorithm and derive the expected convergence rate for denoising, demonstrating the benefit of including multiple measurement vectors. The second approach aims to recover the population covariance matrix from the partially observed sample covariance matrix by motivating its low-rank Toeplitz structure without recovering the signal ensemble. Performance guarantee is derived with a finite number of measurement vectors. The frequencies can be recovered via conventional spectrum estimation methods such as MUSIC from the estimated covariance matrix.  Finally, numerical examples are provided to validate the favorable performance of the proposed algorithms, with comparisons against several existing approaches.
\end{abstract}

\section{Introduction}
Many signal processing applications encounter a signal ensemble where each signal in the ensemble can be represented as a sparse superposition of $r$ complex sinusoids sharing the same frequencies, for example in remote sensing, array processing and super-resolution imaging, and the goal is to recover the set of signals and their corresponding frequencies from a small number of measurements. While there has been a long line of traditional approaches \cite{scharf1991statistical}, \reviseA{Compressed Sensing (CS)} \cite{CanRomTao06,Don2006} has been recently proposed as an efficient way to reduce the number of measurements with provable performance guarantees by promoting the sparsity prior in the reconstruction in a tractable manner. In particular, it is shown that if the frequencies all lie on the DFT grid, the signal of length $n$ can then be recovered exactly \reviseA{using convex optimization} from an order of $r\log n$ randomly selected samples with high probability \cite{candes2007sparsity}, where $r\ll n$. CS has also found many important applications in analog-to-digital conversion \cite{tropp2010beyond,mishali2010theory}, spectrum estimation \cite{tian2007compressed} and hyperspectral imaging \cite{willett2014sparsity}.


%

However, most existing CS theories act as a model selection principle, where the signal is assumed sparse in an a priori basis, and the goal is to identify the activated atoms in the basis. There is a modeling gap, however, from physical signals that are actually composed of a small number of {\em parameterized} atoms with {\em continuous and unknown} parameters determined by the mother nature. An example in this category that garnered much attention is the spectrally-sparse signal, where the signal is composed of a small number of spectral atoms with arbitrary frequencies on the unit interval. 
Performance degeneration of CS algorithms is observed and studied systematically in \cite{chi2011sensitivity,scharf2011sensitivity,pakroohanalysis} when there is an unavoidable basis mismatch between the actual frequencies and the assumed basis. Many subsequent works have been proposed to mitigate the effect of basis mismatch to a great extent \reviseA{(we only cite a partial list\cite{candes2011compressed,fannjiang2012coherence,zhu2011sparsity,fyhn2013compressive,liu2013spectral,duarte2013spectral,ekanadham2011recovery} due to space limits)}.

Therefore, it becomes necessary to develop a {\em parameter estimation} principle, which does not need an a priori basis for reconstruction while still explores the sparsity prior. One recent approach is based on atomic norm minimization \cite{chandrasekaran2012convex}, which provides a general recipe for designing convex solutions to parsimonious model selection. It has been successfully applied to recover a spectrally-sparse signal from a small number of consecutive samples \cite{CandesFernandez2012SR} or randomly selected samples \cite{TangBhaskarShahRecht2012} from the time domain. In particular, Tang et. al. showed that a spectrally-sparse signal can be recovered from an order of $r\log n \log r$ random samples with high probability \reviseA{when the frequencies are at least separated by $4/\left(n-1\right)$} \cite{TangBhaskarShahRecht2012} for line spectra with random amplitudes. This approach is extended to higher dimensional frequencies in \cite{chi2015compressive}. Another approach is proposed in \cite{chen2013spectral,chen2014robust} based on structured matrix completion, where the problem is reformulated into a structured multi-fold Hankel matrix completion inspired by the matrix pencil algorithm \cite{Hua1992}. For this approach, it is shown that an order of \reviseA{$r\log^{4}{n}$} randomly selected samples guarantees perfect recovery with high probability under some mild incoherence conditions and the approach is also amenable to higher-dimensional frequencies. Both approaches allow recovering off-the-grid frequencies at an arbitrary precision from a number of samples much smaller than $n$. We refer interested readers for respective papers for details.

\subsection{Our Contributions and Comparisons to Related Work}
It has been shown in the traditional CS framework that the availability of multiple measurement vectors (MMV) can further improve performance by harnessing the joint sparsity pattern of different signals, also known as {\em group sparsity} \cite{tropp2006algorithms2,tropp2006algorithms,lee2012subspace,kim2012compressive,davies2012rank,mishali2008reduce}. \reviseA{Motivated by recent advances of off-the-grid frequency estimation in the single measurement vector case \cite{CandesFernandez2012SR,chen2013spectral,chen2014robust,TangBhaskarShahRecht2012}, we study the problem of line spectrum estimation and denoising of multiple spectrally-sparse signals from their possibly partial and noisy observations, where all the signals are composed of a common set of continuous-valued frequencies, where we leverage the power of MMV without assuming the frequencies to lie exactly on a grid.}

Two approaches are developed based on atomic norm minimization and structured covariance estimation, both of which can be solved efficiently using Semi-Definite Programming (SDP). We study both their theoretical properties, and provide numerical examples to validate their favorable performance with comparisons to several existing methods, demonstrating the performance gain when the number of measurement vectors increases. 

The first approach can be regarded as a continuous counterpart of the MMV model in CS. \reviseA{Inspired \cite{TangBhaskarShahRecht2012}}, we first define the atomic norm of multiple spectrally-sparse signals and characterize its semidefinite program formulation, \reviseA{which extends the atomic norm for a single spectrally-sparse signal first defined in \cite{TangBhaskarShahRecht2012} to the MMV case}. We then consider signal recovery from their partial noiseless observations, and signal denoising from their full observations in Additive White Gaussian Noise (AWGN), based on atomic norm minimization under the respective observation models. We characterize the dual problem of the proposed algorithm and outline frequency recovery by examining the dual polynomial. \reviseA{In the noiseless case, we show that the same argument in \cite{TangBhaskarShahRecht2012} also leads to a performance guarantee of the MMV case, where we exactly recover the signal ensemble with high probability, as soon as the number of samples per measurement vector is on the order of $r\log n \log r$ under the same separation condition.} In the noisy case, we derive the expected convergence rate for denoising with full observations as a function of the number of measurement vectors, demonstrating the benefit of including MMV. 


A disadvantage of the above approach is that the computational complexity becomes expensive when the number of measurement vectors is high if we wish to recover the whole signal ensemble. Recognizing that in many scenarios one only wishes to recover the set of frequencies, we switch our focus on reconstructing the covariance matrix rather than the signal ensemble in the second approach. Covariance structures can be explored when multiple observations of a stochastic signal are available \cite{tourtier1987maximum}. With a mild second-order statistical assumption on the sparse coefficients, a correlation-aware approach is proposed in \cite{pal2012correlation, pal2012application} to improve the size of recoverable support by exploring the sparse representation of the covariance matrix in the Khatri-Rao product of the signal sparsity basis. However, due to the earlier-mentioned basis mismatch issue, the correlation-aware approach cannot estimate frequencies off the grid.

Under the statistical assumption that the frequencies are uncorrelated which holds in a variety of applications in array signal processing \cite{scharf1991statistical}, the full covariance matrix is a Hermitian Toeplitz matrix whose rank is the number of distinct frequencies. In the second approach, we first calculate the partial sample covariance matrix from partial observations of the measurement vectors. A convex optimization algorithm is formulated to estimate the full Hermitian Toeplitz covariance matrix whose submatrix on the set of observed entries is close to the partial sample covariance matrix, with an additional trace regularization that promotes the Toeplitz low-rank structure. Trace regularization for positive semidefinite matrices is a widely adopted convex relaxation of the non-convex rank constraint. We derive non-asymptotic performance guarantee of the proposed structured covariance estimation algorithm with a finite number of measurement vectors assuming full observations or partial observations using a complete sparse ruler \cite{Shakeri2012ruler}. Finally, the set of frequencies can be obtained from the estimated covariance matrix using conventional methods such as MUSIC \cite{schmidt1986multiple}. Compared with directly applying MUSIC to the partial sample covariance matrix, the proposed algorithm has the potential to recover a higher number of frequencies than the number of samples per measurement vector by taking advantages of the array geometry, for example the co-prime array \cite{pal2011coprime} or the minimum sparse ruler \cite{Shakeri2012ruler}. As this algorithm only requires the partially observed sample covariance matrix rather than the observed signals, the computational complexity does not grow with the number of measurement vectors, in contrast to approaches that aim to recover the signal ensemble. 

We note that several recent papers \cite{yang2014discretization, pal2014gridless} have also proposed discretization-free approaches for direction-of-arrival estimation by exploiting low-rank properties of the covariance matrix under different setups. However, only statistical consistency is established for the algorithm in \cite{yang2014discretization} without a finite sample analysis. The paper \cite{pal2014gridless} assumes completed observation of the covariance matrix and applies low-rank matrix denoising under specific array geometries without performance guarantees. 

\subsection{Paper Organization \reviseA{and Notations}}
The rest of the paper is organized as below. Section~\ref{backgrounds} describes the signal model with MMV and defines its atomic norm. Section~\ref{sec:atomic_denoising} considers line spectrum estimation and denoising based on atomic norm minimization, and Section~\ref{sec:algorithm} presents the second algorithm based on \reviseA{structured covariance estimation}. Numerical experiments are provided in Section \ref{numerical} to validate the proposed algorithms. Finally, conclusions and future work are discussed in Section \ref{sec:conclusion}. Throughout the paper, matrices are denoted by bold capitals and vectors by bold lowercases. The transpose is denoted by $(\cdot)^T$, and the complex conjugate or Hermitian is denoted by $(\cdot)^*$.




\section{Signal Model with MMV and its Atomic Norm} \label{backgrounds}
 In this section we first describe the spectrally-sparse signal model with multiple vectors, then define and characterize the atomic norm associated with the MMV model for spectrally-sparse signals.
 
 \subsection{Signal Model with MMV}
Let $\bx=[x_1,\ldots,x_n]^T\in\mathbb{C}^n$ be a spectrally-sparse signal with $r$ distinct frequency components, written as
\begin{equation}\label{single_mv}
\bx= \sum_{k=1}^r c_{k} \ba(f_k) \triangleq \bV  \bc,
\end{equation}
where each atom $\ba(f)$ is defined as
\begin{equation}
\ba(f ) = \frac{1}{\sqrt{n}}\left[1, e^{j2\pi f}, \ldots, e^{j2\pi f(n-1)}\right]^T , \quad f \in [0,1),
\end{equation}
the matrix $\bV$ is given as $\bV= [\ba(f_1),\ldots, \ba(f_r)]\in\mathbb{C}^{n\times r}$, and $\bc=[c_1,\ldots,c_r]^T\in\mathbb{C}^r$. The set of frequencies $\mathcal{F}=\{f_k\}_{k=1}^r$ can lie anywhere on the unit interval, so that $f_k$ is continuous-valued in $[0,1)$. 

In an MMV model, we consider $L$ signals, stacked in a matrix, $\bX=[\bx_1,\ldots,\bx_L]$, where each signal $\bx_l\in\mathbb{C}^{n}$, $l=1,\ldots, L$, is composed of
\begin{equation}\label{equ:signaldefine}
\bx_l = \sum_{k=1}^r c_{k,l} \ba(f_k) = \bV \bc_l,
\end{equation}
with $\bc_l = [c_{1,l},\ldots,c_{r,l}]^T$. Hence $\bX$ can be expressed as
\begin{equation}\label{data_rep}
\bX = \bV\bC,
 \end{equation}
where $\bC =[\bc_1, \cdots, \bc_L ]\in\mathbb{C}^{r \times L}$.

\subsection{Atomic Norm of the MMV Model} \label{proposal}
We follow the general recipe proposed in \cite{chandrasekaran2012convex} to define the atomic norm of a spectrally-sparse signal ensemble $\bX$. We first define an atom for representing $\bX$ in \eqref{data_rep} as
 \begin{equation}
\bA(f,\bb ) = \ba(f)\bb^* , 
\end{equation}
where $ f \in [0,1)$, $\bb\in\mathbb{C}^{L}$ with $\|\bb\|_2 =1$. The atomic set is defined as $\cA=\{\bA(f,\bb ) | f \in [0,1), \|\bb\|_2 =1\}$. Define
\begin{align}\label{atomic_zero}
\| \bX  \|_{\cA,0} &  = \inf_r \left\{   \bX = \sum_{k=1}^r c_k \bA(f_k,\bb_k), c_k \geq 0 \right\},
\end{align}
as the smallest number of atoms to describe $\bX$. A natural objective to describe $\bX$ is to minimize $\| \bX \|_{\cA,0}$, i.e. to seek the atomic decomposition of $\bX$ with the minimal number of atoms.
It is easy to show that $\| \bX \|_{\cA,0}$ can be represented equivalently as \cite{TangBhaskarShahRecht2012}
\begin{equation}
\begin{split}
\| \bX\|_{\cA,0} &  =  \inf_{\bu ,\bW } \left\{  \rank(\toep(\bu))  \Big| \begin{bmatrix}
\toep(\bu) & \bX\\
\bX^* & \bW \end{bmatrix} \succeq \bf{0} \right\},
\end{split}
\end{equation}
where $\mathcal{T}\left(\boldsymbol{u}\right)$ is the Hermitian Toeplitz matrix with vector $\boldsymbol{u}$ as its first column. Since minimizing \eqref{atomic_zero} is NP-hard, we consider the convex relaxation of $\| \bX\|_{\cA,0}$, called the atomic norm of $\bX$, as
\begin{align}
&\| \bX \|_{\cA} = \inf \left\{ t>0: \; \bX\in t\; \mbox{conv}(\cA) \right\} \nonumber \\
&\quad= \inf \left\{ \sum_k c_k \Big| \bX = \sum_k c_k \bA(f_k,\bb_k), c_k \geq 0 \right\}, \label{atomic_def}
\end{align}
where $\mbox{conv}(\cA)$ is the convex hull of $\cA$. This definition generalizes the atomic norm of a single vector $\bx_l$ in \cite{TangBhaskarShahRecht2012}, which becomes a special case of \eqref{atomic_def} for $L=1$. 



Encouragingly, the atomic norm $\left\Vert\boldsymbol{X}\right\Vert_{\mathcal{A}}$ admits the following equivalent SDP characterization, which implies efficient computation. The proof can be found in Appendix~\ref{proof_atomic_sdp}.

\begin{theorem}\label{atomic-sdp} The atomic norm $\|\bX\|_\cA$ can be written equivalently as
\begin{align*}
 \| \bX\|_\cA = \inf_{\bu\in\mathbb{C}^n,\bW\in\mathbb{C}^{L\times L}} & \Big\{ \frac{1}{2}\trace(\toep(\bu)) + \frac{1}{2}\trace(\bW) \Big|  \\
 & \begin{bmatrix}
\toep(\bu) & \bX \\
\bX^* & \bW \end{bmatrix} \succeq \bf{0} \Big\},
\end{align*}
where $\mbox{Tr}\left(\boldsymbol{X}\right)$ is the trace of $\boldsymbol{X}$.
\end{theorem}

\section{Atomic Norm Miminization With MMV Model}\label{sec:atomic_denoising}

In this section, we consider line spectrum estimation and denoising based on atomic norm minimization from partial and/or noisy observations of multiple spectral-sparse signals: (a) signal recovery from their partial noiseless observations; and (b) denoising from their full observations in AWGN.


We assume that a random or deterministic (sub)set of entries of each vector in $\boldsymbol{X}^{\star}$ defined in \eqref{data_rep} is observed, and the observation pattern is denoted by $\bar{\Omega}\subset\{0,\ldots, n-1\}\times \{1, \ldots, L\}$. In the absence of noise, the partially observed signal matrix is given as
\begin{equation}
\boldsymbol{Z}_{\bar{\Omega}}=\boldsymbol{X}_{\bar{\Omega}}^{\star}=\mathcal{P}_{\bar{\Omega}}(\boldsymbol{X}^{\star}), 
\end{equation}
where $\mathcal{P}_{ \bar{\Omega}}$ is a projection matrix on the set indexed by $ \bar{\Omega}$. Note that we allow the observation pattern of each column of $\bX^{\star}$ to be different, possibly randomly selected. 

We propose the following atomic norm minimization algorithm to recover the complete signal $\bX^{\star}$:
\begin{equation}\label{primal}
\hat{\bX}=\argmin_\bX \|\bX\|_\cA \quad \mbox{s.t.} \quad \bX_{\bar{\Omega}} = \bZ_{\bar{\Omega}}.
\end{equation}
When the measurements are corrupted by noise, give as
$$\bZ_{\bar{\Omega}}=\bX_{\bar{\Omega}}^{\star}+\bN_{\bar{\Omega}},$$
where $\bN_{\bar{\Omega}}$ is the noise term, we consider the atomic norm regularized algorithm:
\begin{equation} \label{primal_noisy}
\hat{\bX}=\argmin_\bX \; \frac{1}{2}  \|\bX_{\bar{\Omega}} - \bZ_{\bar\Omega}\|_F^2 +\tau \|\bX\|_\cA,
\end{equation}
where $\tau$ is a carefully-selected regularization parameter \reviseA{(c.f. \cite{Bhaskar2013denoising})}. We will first analyze the noiseless algorithm \eqref{primal} with partial observations in Section~\ref{sec:partial_recovery} and then the denoisng algorithm \eqref{primal_noisy} will full observations in Section~\ref{sec:denoising}. The theoretical analysis of the case with partial noisy observations is left to future work.

\subsection{Signal Recovery from Partial Noiseless Observations}\label{sec:partial_recovery}
From Theorem~\ref{atomic-sdp}, we can equivalently write \eqref{primal} as the following semidefinite program:
\begin{align}
\hat{\bX}= \argmin_{\bX}\inf_{\bu,\bW} &\;  \frac{1}{2}\trace(\toep(\bu)) + \frac{1}{2}\trace(\bW) \label{primal-sdp}\\
 \mbox{s.t.} &\;  \begin{bmatrix}
\toep(\bu) & \bX \\
\bX^* & \bW \end{bmatrix} \succeq \mathbf{0} , \bX_{\bar\Omega} = \bZ_{\bar\Omega}. \nonumber
\end{align}
Similarly, \eqref{primal_noisy} can be recast as a semidefinite program as well.



Interestingly, one can recover the set of frequencies from the solution of the dual problem of \eqref{primal}. Define $\langle \bY,\bX \rangle = \trace(\bX^*\bY)$, and $\langle \bY,\bX \rangle_{\mathbb{R}}=\mbox{Re}(\langle \bY,\bX \rangle)$. The dual norm of $\|\bX\|_\cA$ can be defined as
\begin{align*}
\|\bY\|_{\cA}^* &=\sup_{\|\bX\|_\cA\leq 1} \langle \bY,\bX\rangle_{\mathbb{R}} \\
&  = \sup_{f\in[0,1),\|\bb\|=1}  \langle \bY, \ba(f) \bb^* \rangle_{\mathbb{R}} \\
&  = \sup_{f\in[0,1),\|\bb\|=1} \left| \langle \bb, \bY^*\ba(f)   \rangle \right| \\
& = \sup_{f\in[0,1)} \|\bY^*\ba(f)  \|_2 = \sup_{f\in[0,1)} \| \bQ(f)\|_2.
\end{align*}

The dual problem of \eqref{primal} can be written as
\begin{equation} \label{dual}
\hat{\bY} = \argmax_{\bY} \; \langle \bY_{\bar\Omega}, \bZ_{\bar\Omega} \rangle_{\mathbb{R}} \; \mbox{s.t.} \; \|\bY\|_{\cA}^* \leq 1, \bY_{{\bar\Omega}^c} = 0.
\end{equation}
Following \cite{CandesFernandez2012SR,chen2013spectral,chen2014robust,TangBhaskarShahRecht2012}, one can recover the set of frequencies using a dual polynomial $\|\bQ(f)\|_2=\| \hat{\boldsymbol{Y}}^* \boldsymbol{a}(f) \|_2$ constructed from the dual solution $\hat{\boldsymbol{Y}}$, by identifying the frequencies that satisfy $\{f\in [0,1): \|\bQ(f)\|_2 = 1\}$. Once the frequencies are identified, their amplitudes can be recovered by solving a follow-up group sparsity minimization problem.

Let $(\bX,\bY)$ be primal-dual feasible to \eqref{primal} and \eqref{dual}, we have $\langle \bY,\bX\rangle_{\mathbb{R}}=\langle \bY,\bX^{\star}\rangle_{\mathbb{R}}$. Strong duality holds since Slater's condition holds \cite[Chapter 5]{boyd2004convex}, and it implies that the solutions of \eqref{primal} and \eqref{dual} equal if and only if $\bY$ is dual optimal and $\bX$ is primal optimal. Using strong duality, we have the following proposition to certify the optimality of the solution of \eqref{primal} whose proof can be found in Appendix~\ref{proof_dual_certificate}.

\begin{prop} \label{dual_certificate}The solution of \eqref{primal} $\hat{\bX}=\bX^{\star}$ is its unique optimizer if there exists $\bY$ such that $\bY_{{\bar\Omega}^c} = 0$ and $\bQ(f)= \bY^*\ba(f)$ satisfies
\begin{equation} \label{conditions}
\begin{cases}
  \bQ(f_k) = \bb_k, & \forall f_k \in \mathcal{F}, \\
  \| \bQ(f) \|_2 < 1, &\forall f\notin \mathcal{F}.
\end{cases}
\end{equation}

\end{prop}


Proposition~\ref{dual_certificate} offers a way to certify the optimality of \eqref{primal} as long as we can find a dual polynomial $\bQ(f)$ that satisfies \eqref{conditions}. Borrowing the dual polynomials constructed for the single measurement vector case in \cite{TangBhaskarShahRecht2012}, we can easily show that the atomic norm minimization for MMV models succeeds with high probability under the same frequency separation condition when the size of $\bar{\Omega}$ exceeds certain threshold. We have the following theorem.
 
\begin{theorem}\label{mmv_atomic_guarantee}
Let $\bar{\Omega}$ be a set of indices selected uniformly at random from $\{0,\ldots, n-1\}\times \{1, \ldots, L\}$. Additionally, assume the signs $c_{k,l}/|c_{k,l}|$ are drawn i.i.d. from the uniform distribution on the complex unit circle and that
\begin{equation}\label{separation}
 \Delta : = \min_{k\neq l}  |f_k -f_l | \geq \frac{1}{\lfloor (n-1)/4 \rfloor}
 \end{equation}
which is the minimum separation between frequency pairs wrapped around on the unit circle. Then there exists a numerical constant $C$ such that
\begin{equation}\label{sample_complexity}
|  \bar{\Omega} | \geq C L \max\left \{ \log^2\frac{n}{\delta}, r\log\frac{r}{\delta} \log\frac{n}{\delta}\right\}  
\end{equation}
is sufficient to guarantee that we can recover $\bX$ via \eqref{primal} with probability at least $1-L\delta$.
\end{theorem}

From Theorem~\ref{mmv_atomic_guarantee} we can see that the atomic norm minimization succeeds with high probability as soon as the number of samples is slightly above the information-theoretical lower bound $\Theta(rL)$ by logarithmic factors, given a mild separation condition is satisfied. Theorem~\ref{mmv_atomic_guarantee} is a straightforward extension of the single vector case $L=1$ studied in \cite{TangBhaskarShahRecht2012}, by constructing each row of $\bQ(f)$ in the same manner as \cite{TangBhaskarShahRecht2012}, hence the proof is omitted. On average, the number of samples per measurement vector is about $|\bar{\Omega}|/L$, which is on the order of $r\log n \log r$, similar to the single vector case \cite{TangBhaskarShahRecht2012}. Nonetheless, we demonstrate in the numerical examples in Section~\ref{numerical} that indeed the inclusion of multiple vectors can improve the reconstruction performance. Therefore, it will be interesting to see whether one can relax either \eqref{separation} or \eqref{sample_complexity} given more measurement vectors.\footnote{A recent preprint \cite{yang2014exact} appeared on Arxiv while this work was under preparation slightly improves the probability of success of Theorem~\ref{mmv_atomic_guarantee} from $1-L\delta$ to $1-\sqrt{L}\delta$ using more refined arguments.}

\begin{remark} (Connection to the single vector case)
It is possible to employ the atomic norm minimization for the MMV model to recover a partially observed spectrally-sparse signal. Specifically, consider a Hankel matrix constructed from $\bx$ in \eqref{single_mv} as
\begin{equation} \label{hankel}
\mathcal{H}(\bx, p)  = \begin{bmatrix}
x_1 & x_2 & \cdots & x_{n-p+1}\\
x_2 & x_3 & \cdots & x_{n-p+2} \\
\vdots & \vdots & \ddots & \vdots \\
x_p & x_{p+1} & \cdots & x_n
\end{bmatrix},
\end{equation}
where $p$ is a pencil parameter. We can then view the columns of $\mathcal{H}(\bx,p)$ as an ensemble of spectrally-sparse signals sharing the same frequencies. We may propose to minimize the atomic norm of $\mathcal{H}(\bx, p)$ as
\begin{equation}\label{atomic_hankel}
\hat{\bx}_{\cA}  = \argmin_{\bx}\; \|\mathcal{H}(\bx, p)  \|_{\cA} \quad \mbox{s.t. } \quad  \bx_{\Omega} = \bz_{\Omega},
\end{equation}
which can be reformulated as
\begin{align}
& \min_{\bu,\bW_2,\bx} \trace(\toep(\bu)) + \trace(\bW_2) \label{atomic_reformulation} \\
&\quad \mbox{s.t. }\begin{bmatrix}
\toep(\bu) & \mathcal{H}(\bx, p) \\
\mathcal{H}(\bx, p)^* & \bW_2 
\end{bmatrix} \succeq 0,\;  \bx_{\Omega} = \bz_{\Omega}. \nonumber
\end{align}

This draws an interesting connection to the Enhanced Matrix Completion (EMaC) algorithm proposed in \cite{chen2013spectral,chen2014robust}, which recovers $\bx$ by minimizing the nuclear norm of $\mathcal{H}(\bx, p)$ as
\begin{equation} \label{nuclear_hankel}
\hat{\bx}_{\text{EMaC}} =  \argmin_{\bx}\; \|\mathcal{H}(\bx, p)  \|_{*} \quad \mbox{s.t. }   \bx_{\Omega} = \bz_{\Omega},
\end{equation}
which can be reformulated as
\begin{align}
&\min_{\bW_1,\bW_2,\bx} \trace(\bW_1) + \trace(\bW_2) \label{nuclear_reformulation} \\
&\quad \mbox{s.t. }\begin{bmatrix}
\bW_1 & \mathcal{H}(\bx, p) \\
\mathcal{H}(\bx, p)^* & \bW_2 
\end{bmatrix} \succeq 0,\;  \bx_{\Omega} = \bz_{\Omega}. \nonumber 
\end{align}

\reviseA{Comparing \eqref{atomic_reformulation} and \eqref{nuclear_reformulation}, the EMaC algorithm can be regarded as a relaxation of \eqref{atomic_reformulation} by dropping the Toeplitz constraint (which allows handling of damping modes) of the first diagonal block in EMaC.  When $p=1$, \eqref{atomic_reformulation} is equivalent to the atomic norm minimization algorithm in \cite{TangBhaskarShahRecht2012}. Note that Theorem~\ref{mmv_atomic_guarantee} cannot be applied to guarantee the success of \eqref{atomic_reformulation} since the signs of each vector are not independent, and in practice this formulation does not provide performance gains over the atomic norm minimization algorithm in \cite{TangBhaskarShahRecht2012}. Hence we present this formulation just for theoretical interests. }

\end{remark}

\subsection{Signal Denoising for MMV Model}\label{sec:denoising}
In this section, we consider the problem of line spectrum denoising in AWGN when full observations are available. The algorithm \eqref{primal_noisy} can be rewritten as
\begin{equation} \label{equ:equivalent}
\hat{\boldsymbol{X}}=\argmin_{\boldsymbol{X}}\frac{1}{2} \left\Vert\boldsymbol{X}-\boldsymbol{Z} \right\Vert_{F}^{2}+\tau\left\Vert\boldsymbol{X}\right\Vert_{\mathcal{A}},
\end{equation}
where the subscript $\bar{\Omega}$ is dropped with $\bZ = \bX^{\star} + \bN $ and $\bN$ is the additive noise. This algorithm can be efficiently implemented via ADMM, of which we provide the procedure in Appendix~\ref{ADMM}. We have the following theorem for the expected convergence rate of \eqref{equ:equivalent} when the noise is AWGN. The proof is in Appendix~\ref{proof_mmv_rate}.

\begin{theorem}\label{thm_mmv_rate}
Assume the entries of $\boldsymbol{N}$ are composed of i.i.d. Gaussian entries $\mathcal{CN}(0,\sigma^2)$. Set $\tau = \sigma  \left(1+\frac{1}{\log{n}}\right)^{\frac{1}{2}}  \left(L+\log{\left(\alpha L\right)}+\sqrt{2L\log{\left(\alpha L\right)}}+\sqrt{\frac{\pi L}{2}}+1\right)^{\frac{1}{2}}$, where $\alpha=8\pi n\log{n}$, then the expected convergence rate is bounded as
\begin{equation}\label{mmv_rate}
\frac{1}{L}\mathbb{E} \left\Vert\hat{\boldsymbol{X}}-\boldsymbol{X}^{\star}\right\Vert_{\mathrm{F}}^{2}\leq \frac{2\tau}{L} \left\Vert\boldsymbol{X}^{\star}\right\Vert_{\mathcal{A}}.
\end{equation}
\end{theorem}

From Theorem~\ref{thm_mmv_rate},  $\tau$ is set on the order of $\sqrt{L}$. If $\left\Vert\boldsymbol{X}^{\star}\right\Vert_{\mathcal{A}}=o\left(\sqrt{L}\right)$, then the per-measurement vector Mean Squared Error (MSE) vanishes as $L$ increases. This is satisfied, for example by a correlated signal ensemble where the norm of each row of coefficient matrix $\bC$ is $o(\sqrt{L})$. On the other hand, if all entries in $\bC$ are selected with unit amplitude, then $\left\Vert\boldsymbol{X}^{\star}\right\Vert_{\mathcal{A}}= O(\sqrt{L})$ and the per-measurement vector MSE may not vanish with the increase of $L$. Nonetheless, our numerical examples in Section~\ref{sec:denoising_admm} demonstrate that the per-measurement vector MSE decreases gracefully with the increase of $L$.

\section{Structured Covariance Estimation for MMV Model}\label{sec:algorithm}
While the availability of MMV improves the performance as demonstrated in Section~\ref{sec:atomic_denoising}, the computational cost also increases dramatically when $L$ is large. In many applications, one is only interested in the set of frequencies, and the covariance matrix of the signal carries sufficient information to recover the frequencies. In this section, we develop a {structured} covariance estimation algorithm that takes advantages of statistical properties of the frequency coefficients and the low-dimensional structures of the covariance matrix. 

In particular, we assume that the coefficients $c_{k,l}$'s satisfy $\mathbb{E}[c_{k,l}]=0$ and the following second-order statistical property:
 \begin{equation}\label{second_order}
 \mathbb{E}[c_{k,l} c_{k^{\prime},l^{\prime}}] =\left\{ \begin{array}{cc}
 \sigma_k^2, & \mbox{if}~ k = k^{\prime},\; l = l^{\prime}, \\
 0, & \mbox{otherwise}.
 \end{array}\right.
 \end{equation}
To put it differently, the coefficients from different signals are uncorrelated, and the coefficients for different frequencies in the same signal are also uncorrelated. As an example, \eqref{second_order} is satisfied if $c_{k,l}$'s are generated i.i.d. from $\mathcal{CN}\left(0,\sigma_k^{2}\right)$. 

Assume each vector in $\bX $ is observed at the same location $\Omega$ of size $m$. Without ambiguity, we use $\Omega$ to denote both the observation pattern of the signal ensemble $\bX_\Omega$ and each signal $\bx_{\Omega,l}$. Instead of focusing on reconstructing the complete signal matrix $\bX$, we explore the low-dimensional structure of its covariance matrix. Given \eqref{second_order}, it is straightforward that the covariance matrix of the signal $\boldsymbol{x}_{l}$ in \eqref{equ:signaldefine} can be written as
\begin{equation}\label{equ:covarianceidealfull}
\boldsymbol{\Sigma}^{\star}=\mathbb{E}\left[\boldsymbol{x}_{l}\boldsymbol{x}_{l}^{*}\right]=\sum_{k=1}^r\sigma_k^{2}\boldsymbol{a}\left(f_k\right)\boldsymbol{a}\left(f_k \right)^{*}\reviseA{\in\mathbb{C}^{n\times n}},
\end{equation}
which is a Positive Semi-Definite (PSD) Hermitian Toeplitz matrix. This matrix is low-rank with $\mbox{rank}\left(\boldsymbol{\Sigma}^{\star}\right)=r\ll n$. In other words, the spectral sparsity translates into the small rank of the covariance matrix. Let the first column of $\boldsymbol{\Sigma}^{\star}$ be $\boldsymbol{u}^\star=\frac{1}{\sqrt{n}}\sum_{k=1}^{r}\sigma_{k}^{2}\boldsymbol{a}\left(f_{k}\right)\in\mathbb{C}^{n}$, then $\boldsymbol{\Sigma}^{\star}$ can be rewritten as $\boldsymbol{\Sigma}^{\star} = \mathcal{T}\left(\boldsymbol{u}^\star \right)$. From $\boldsymbol{u}^\star$ or $\boldsymbol{\Sigma}^{\star}$, the set of frequencies can be estimated accurately by well-studied spectrum estimation algorithms such as MUSIC \cite{schmidt1986multiple} and ESPRIT \cite{roy1989esprit}. Therefore, we focus ourselves on reconstruction of the covariance matrix $\boldsymbol{\Sigma}^{\star}$.

\subsection{Structured Covariance Estimation with SDP}
The covariance matrix of the partially observed samples $\boldsymbol{x}_{\Omega,l}$ can be given as
\begin{equation}\label{equ:covarianceideal}
\boldsymbol{\Sigma}^{\star}_{\Omega} = \mathbb{E}[\boldsymbol{x}_{\Omega,l}\boldsymbol{x}_{\Omega,l}^*] = \mathcal{P}_\Omega(\boldsymbol{\Sigma}^{\star}) \in \mathbb{C}^{m\times m},
\end{equation}
where $\mathcal{P}_{\Omega}$ is a mask operator that only preserves the submatrix in the rows and columns indexed by $\Omega$. 

If $\boldsymbol{\Sigma}^{\star}_{\Omega}$ can be perfectly estimated, e.g. using an infinite number of measurement vectors, one might directly seek a low-rank Hermitian Toeplitz matrix $\mathcal{T}(\boldsymbol{u})$ which agrees with $\boldsymbol{\Sigma}^{\star}_{\Omega}$ restricted on the submatrix indexed by $\Omega$. Unfortunately, the ideal covariance matrix in \eqref{equ:covarianceideal} cannot be perfectly obtained; rather, we will first construct the sample covariance matrix of the partially observed samples as
\begin{equation}\label{equ:covariancesample}
\boldsymbol{\Sigma}_{\Omega,L}=\frac{1}{L}\sum_{l=1}^{L}\boldsymbol{x}_{\Omega,l}\boldsymbol{x}_{\Omega,l}^{*}=\frac{1}{L}\boldsymbol{X}_{\Omega}\boldsymbol{X}_{\Omega}^{*}\in\mathbb{C}^{m\times m}.
\end{equation}
Further denote the sample covariance matrix as $\boldsymbol{\Sigma}_{L} = \frac{1}{L }\sum_{l=1}^{L}\boldsymbol{x}_{l}\boldsymbol{x}_{l}^{*}$. We then seek a low-rank PSD Hermitian Toeplitz matrix whose restriction on the submatrix indexed by $\Omega$ is close to the sample covariance matrix $\boldsymbol{\Sigma}_{\Omega,L}$ in \eqref{equ:covariancesample}. A natural algorithm would be
\begin{align}
 \hat{\boldsymbol{u}} & = \argmin_{\boldsymbol{u}\in \mathbb{C}^{n}} \;  \frac{1}{2}\left\Vert\mathcal{P}_{\Omega} \left( \mathcal{T}\left(\boldsymbol{u}\right) \right)-\boldsymbol{\Sigma}_{\Omega,L} \right\Vert_{F}^{2} + \lambda \mbox{rank}\left(\mathcal{T}\left(\boldsymbol{u}\right)\right) \nonumber \\
& \mbox{s.t.} \quad   \mathcal{T}\left(\boldsymbol{u}\right)\succeq 0, \label{eq:rankmin}
\end{align}
where $\lambda$ is a regularization parameter balancing the data fitting term and the rank regularization term. However, as directly minimizing the rank is NP-hard, we consider a convex relaxation for rank minimization over the PSD cone, which replaces the rank minimization by trace minimization, resulting in 
\begin{align}
 \hat{\boldsymbol{u}} &= \argmin_{\boldsymbol{u}\in \mathbb{C}^{n}} \;   \frac{1}{2}\left\Vert \mathcal{P}_{\Omega} \left( \mathcal{T}\left(\boldsymbol{u}\right) \right)-\boldsymbol{\Sigma}_{\Omega,L} \right\Vert_{F}^{2} + \lambda \mbox{Tr}\left(\mathcal{T}\left(\boldsymbol{u}\right)\right)  \nonumber \\
& \mbox{s.t.} \quad      \mathcal{T}\left(\boldsymbol{u}\right)\succeq 0.\label{equ:optimization}
\end{align}
The algorithm \eqref{equ:optimization} can be solved efficiently using off-the-shelf semidefinite program solvers. Interestingly, the trace minimization of $\toep(\bu)$ is equivalent to minimizing the atomic norm of $\boldsymbol{u}$ under the nonnegative constraint $\toep(\bu) \succeq 0$ since $\| \boldsymbol{u}  \|_{\cA}  = \trace(\toep(\bu))$ if $\toep(\bu) \succeq 0$. Therefore we can \reviseA{equivalently write} \eqref{equ:optimization} as an atomic norm regularized algorithm: 
\begin{align*}
 \hat{\boldsymbol{u}} &= \argmin_{\boldsymbol{u}\in \mathbb{C}^{n}} \;   \frac{1}{2}\left\Vert \mathcal{P}_{\Omega} \left( \mathcal{T}\left(\boldsymbol{u}\right) \right)-\boldsymbol{\Sigma}_{\Omega,L} \right\Vert_{F}^{2} + \lambda \| \boldsymbol{u}  \|_{\cA}   \\
& \mbox{s.t.} \quad      \mathcal{T}\left(\boldsymbol{u}\right)\succeq 0.
\end{align*}


The proposed algorithm works with the sample covariance matrix $\boldsymbol{\Sigma}_{\Omega,L}$ rather than $\boldsymbol{X}_{\Omega}$ directly. Therefore, it does not require storing $\boldsymbol{X}_{\Omega}$ of size $mL$, but only $\boldsymbol{\Sigma}_{\Omega,L}$ of size $m^2$, which greatly reduces the storage space when $m\ll L$ and may be updated online if the measurement vectors arrive sequentially. 

It is also worthwhile to compare the proposed algorithm \eqref{equ:optimization} with the correlation-aware method in \cite{pal2012correlation, pal2012application}. The method in \cite{pal2012correlation, pal2012application}, when specialized to a unitary linear array, can be regarded as a discretized version of our algorithm (\ref{equ:optimization}), where the atoms $\boldsymbol{a}(f_k)$'s in the covariance matrix (\ref{equ:covarianceideal}) are discretized over a discrete grid. Further numerical comparisons are carried out in Section~\ref{numerical}.

\subsection{Performance Guarantees with Finite Samples}
We analyze the performance of \eqref{equ:optimization} under an additional Gaussian assumption, where each $\boldsymbol{c}_l$ is i.i.d. generated as $\boldsymbol{c}_l\sim\mathcal{CN}(\bf{0},\boldsymbol{\Lambda})$, and therefore $\bx_l \sim \mathcal{CN}(\bf{0},\boldsymbol{\Sigma}^{\star})$. Define the {\em effective rank} of a matrix $\bSigma$ as $r_{\text{eff}}(\bSigma) =  \mbox{Tr}(\bSigma)/\| \bSigma \|$ which is strictly smaller than $r$ and allows the signal to be approximately sparse. We have the following theorem.
\begin{theorem}\label{thm_mmv_cov}
Suppose that $\boldsymbol{c}_l\sim\mathcal{CN}(\bf{0},\boldsymbol{\Lambda})$. Let $\bu^{\star}$ be the ground truth. Set 
$$\lambda \ge C  \max\left\{ \sqrt{\frac{r_{\text{eff}}(\bSigma_{\Omega}^{\star}) \log (Ln)}{L}}, \frac{r_{\text{eff}}(\bSigma_{\Omega}^{\star}) \log (Ln)}{L} \right\} \| \bSigma_{\Omega}^{\star} \| $$ for some constant $C$, then with probability at least $1-L^{-1}$, the solution to \eqref{equ:optimization} satisfies
$$ \| \toep(\hat{\bu} - \bu^\star)   \|_F \leq 16\lambda \sqrt{r}$$
if $\Omega$ corresponds to full observation; and
\begin{equation*}
\frac{1}{\sqrt{n}} \| \hat{\bu} - \bu^\star \|_F \leq 16\lambda \sqrt{r} 
\end{equation*}
if $\Omega$ is a complete sparse ruler such that the unobserved entries can be deduced from differences of observed ones. 
\end{theorem}

The proof is in Appendix~\ref{proof_mmv_cov}. Note that the observation set $\Omega$ is assumed deterministic in Theorem~\ref{thm_mmv_cov}. When full observations are available, our algorithm yields reliable estimate of the covariance matrix as soon as the number of measurement vectors $L$ is on the order of $r_{\text{eff}}(\bSigma^{\star}) r  \log n \leq r^{2}\log n$, which is much smaller than the ambient dimension $n$. When $\Omega$ forms a complete sparse ruler, the average per-entry MSE vanishes as soon as $L$ is on the order of $r_{\text{eff}}(\bSigma_\Omega^{\star}) r \log n \leq r^{2}\log n$.

\section{Numerical Experiments} \label{numerical}

In this section, we evaluate the performance of the proposed algorithms \eqref{primal}, \eqref{equ:equivalent} and \eqref{equ:optimization}. In particular, we examine the influence of the number of measurement vectors and the number of samples per signal on the performance of frequency estimation, and compare the proposed algorithms against several existing approaches.


\subsection{Atomic Norm Minimization \eqref{primal} for MMV Model  }
\begin{figure}[htp]
\centering
\includegraphics[width=0.4\textwidth]{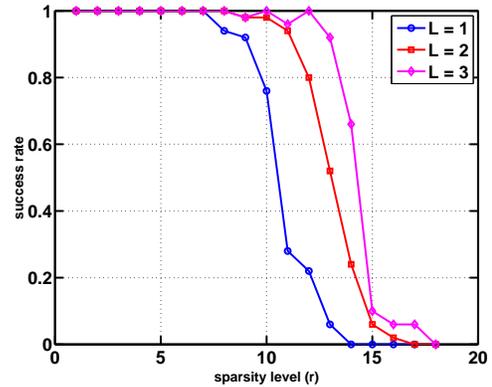}
\caption{Success rate of reconstruction versus the sparsity level $r$ for $L=1,2,3$ when $n=64$, $m=32$ and the frequencies are generated satisfying a separation condition $\Delta\geq1/n$ for the same observation across signals.} \label{fixed_m}
\end{figure}

Let $n=64$ and $m=32$. In each Monte Carlo experiment, we generate $L$ spectrally-sparse signals with $r$ frequencies randomly located in $[0,1)$ that satisfy a separation condition $\Delta =\min_{k\neq l} |f_k-f_l| \geq 1/n$. This separation condition is about $4$ times weaker than the condition asserted in Theorem~\ref{mmv_atomic_guarantee} to guarantee the success of \eqref{primal} with high probability. For each frequency component, we randomly generate its amplitudes for each signal. We run \eqref{primal} and calculate the normalized reconstruction error as $\|\hat{\bX}-\bX^{\star} \|_F/\|\bX^{\star}\|_F$, and claim the experiment is successful if it is below $10^{-5}$. For each pair of $r$ and $L$, we run a total of $50$ Monte Carlo experiments and output the average success rate. Fig.~\ref{fixed_m} shows the success rate of reconstruction versus the sparsity level $r$ for $L=1$, $2$, and $3$ respectively. As we increase $L$, the success rate becomes higher for the same sparsity level.

\begin{figure}[htp]
\begin{center}
\begin{tabular}{c}
\hspace{-0.15in}\includegraphics[height= 1.5in, width=0.45\textwidth]{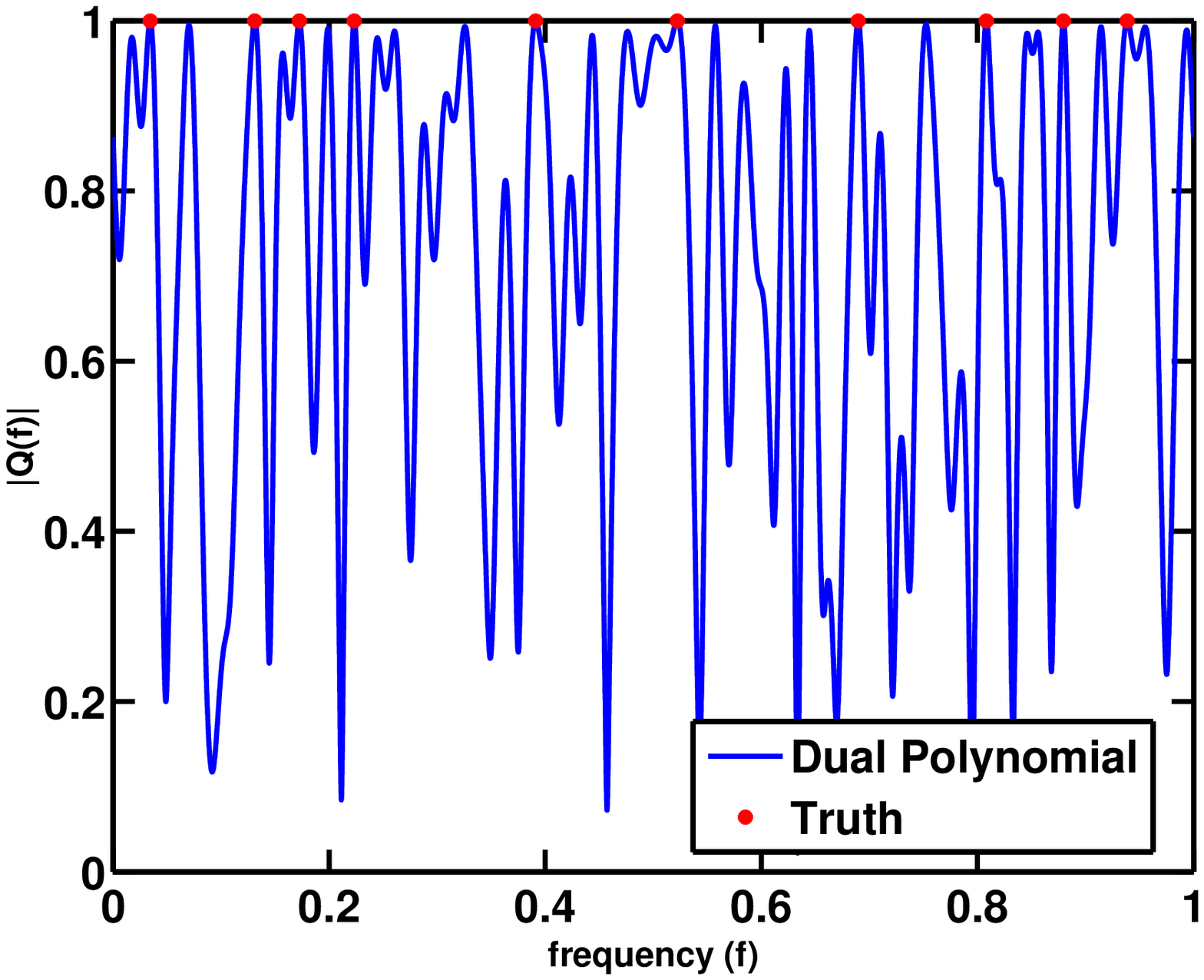} \\
{\scriptsize (a) $  L=1$} \\
\hspace{-0.15in} \includegraphics[height= 1.5in, width=0.45\textwidth]{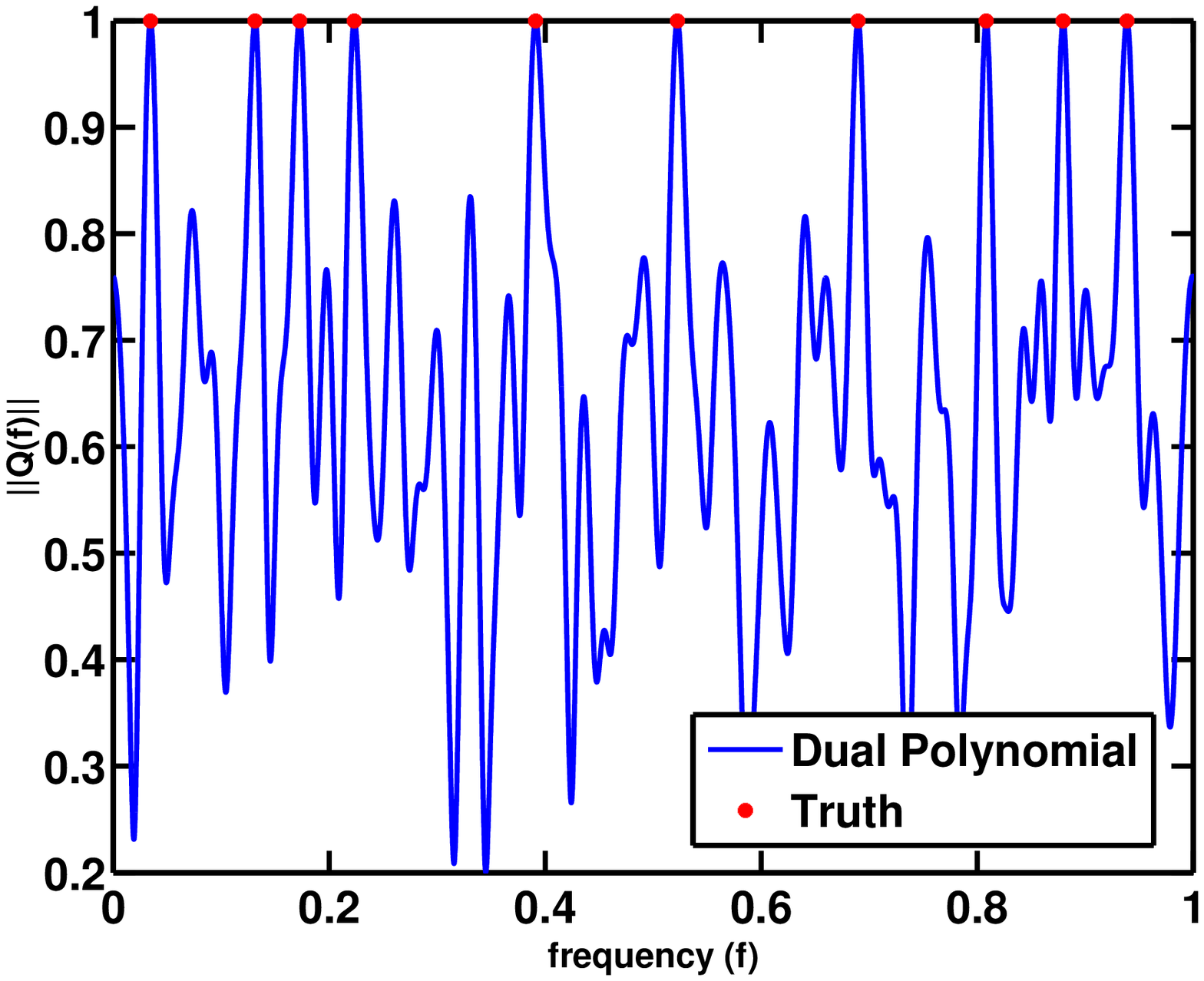}  \\
 {\scriptsize  (b) $  L=3$  }
\end{tabular}
\end{center}
\caption{The reconstructed dual polynomial for a randomly generated spectrally-sparse signal with $n=64$, $r=10$, and $m=32$: (a) $L=1$, (b) $L=3$.} \label{dualpoly}
\end{figure}

Fig.~\ref{dualpoly} shows the reconstructed dual polynomial for a randomly generated spectrally-sparse signal with $r=10$. The amplitudes are generated randomly with $\mathcal{CN}(0,1)$ entries when no noise is present. It can be seen that although the algorithm achieves perfect recovery with both $L=1$ and $L=3$, the reconstructed dual polynomial has a much better localization property when $L=3$.

\subsection{Atomic Norm based Denoising \eqref{equ:equivalent} for MMV Model}\label{sec:denoising_admm}
\begin{figure}[ht]
\centering
\hspace{-0.1in}\includegraphics[height=2.2in, width=0.45\textwidth]{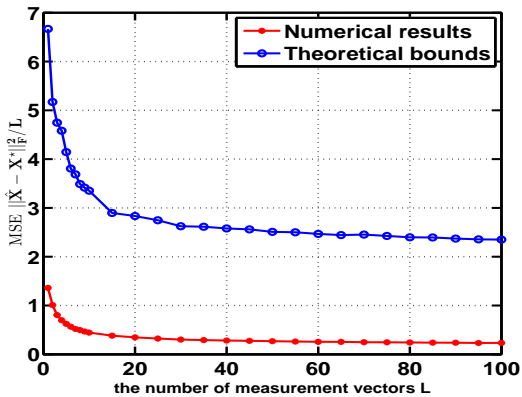}
\caption{The per-measurement vector MSE of reconstruction, and its theoretical upper bound, versus the number of measurement vectors $L$ when $n=64$, $r=8$ and $\sigma = 0.1$.} \label{denoising_admm}
\end{figure}

Let $n=64$ and the sparsity level $r=8$. The frequencies are selected to satisfy the separation condition $\Delta\geq 1/n$. We generate the coefficient matrix $\boldsymbol{C}$ with $c_{k,l}\sim\mathcal{CN}\left(0,1\right)$. The noise matrix $\boldsymbol{N}$ is randomly generated with $\mathcal{CN}\left(0,\sigma^{2}\right)$, where $\sigma=0.1$. We solve \eqref{equ:equivalent} via ADMM and calculate per-measurement vector MSE as $\Vert\hat{\boldsymbol{X}}-\boldsymbol{X}^{\star} \Vert^{2}_{F}/L$. Fig.~\ref{denoising_admm} shows the per-measurement vector MSE of the reconstruction with respect to the number of measurement vectors, together with the theoretical upper bound obtained from Theorem~\ref{thm_mmv_rate}. The per-measurement vector MSE decreases with increasing of $L$, which demonstrates more accurate denoising results brought by MMV. While the theoretical bound is not as tight, it exhibits similar trends as the empirical performance.

\begin{figure}[ht]
\centering
\includegraphics[height=2.4in, width=0.45\textwidth]{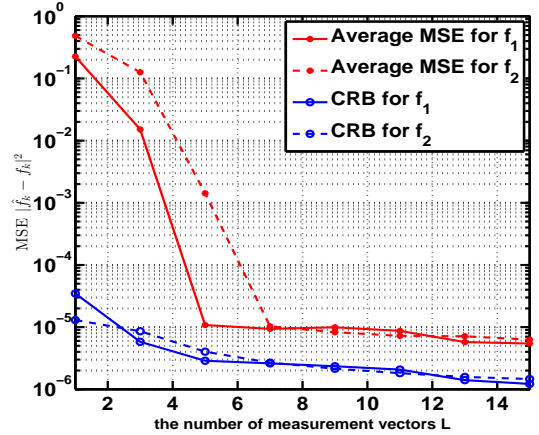}
\caption{The comparison between average MSE of frequency estimates and CRB with respect to $L$ when $n=14$, $r=2$ and $\sigma=0.3$.} \label{f_est_crb}
\end{figure}

We further examine the influence of $L$ on the accuracy of frequency estimation with comparison against the Cramér Rao Bound (CRB). Let $n=14$ and $r=2$. The coefficients $\boldsymbol{C}$ is generated with i.i.d. $\mathcal{CN}\left(0,1\right)$ entries, and the noise is generated with i.i.d. $\mathcal{CN}\left(0,\sigma^{2}\right)$ entries, where $\sigma=0.3$. For each $L$, we obtain the frequency estimates from the dual solution of \eqref{equ:equivalent}, and calculate the MSE of each frequency estimate as $\left(\hat{f_{k}}-f_{k}\right)^{2}$, where $\hat{f_{k}}$ is the estimate of real frequency $f_{k}$, averaged over 500 Monte Carlo runs with respect to the noise realizations. We compare this against the CRB, which can be derived from the following Fisher information matrix $\boldsymbol{J}\left(\boldsymbol{f}\right)$ assuming fixed coefficients:
\begin{align*}
&\boldsymbol{J}\left(\boldsymbol{f}\right)=\frac{8\pi^{2}}{n\sigma^{2}}\sum_{l=1}^{L}\mathrm{Re}\\
&\begin{bmatrix} \left\vert c_{1,l}\right\vert^{2}\sum_{i=0}^{n-1}i^{2} & c_{1,l} c_{2,l}^*\sum_{i=0}^{n-1} i^{2}e^{j2\pi (f_{1}-f_{2})i} \\ c_{1,l}^*c_{2,l}\sum_{i=0}^{n-1}i^{2}e^{j2\pi (f_{2}-f_{1})i} & \left\vert c_{2,l}\right\vert^{2}\sum_{i=0}^{n-1}i^{2} \end{bmatrix}.
\end{align*} 
Fig.~\ref{f_est_crb} shows the average MSE and the corresponding CRB with respect to the number of measurement vectors $L$. With the increase of $L$, the average MSE of frequency estimates approaches to CRB while CRB approaches to $0$.

\subsection{Structured Covariance Estimation \eqref{equ:optimization} for MMV Model}
We conduct several numerical experiments to validate \eqref{equ:optimization}. In particular, we examine the influence of the number of measurement vectors $L$ on the performance of covariance estimation and frequency estimation. Unfortunately we can not directly use Theorem~\ref{thm_mmv_cov} to set $\lambda$ since $\bSigma^{\star}$ is not known. In all the experiments, we set $\lambda=2.5\times 10^{-3}/\left(\left(\log{L}\right)^{2}\log{m}\right)$ which gives good performance empirically. 

\begin{figure}[htp]
\centering
\includegraphics[width=0.45\textwidth]{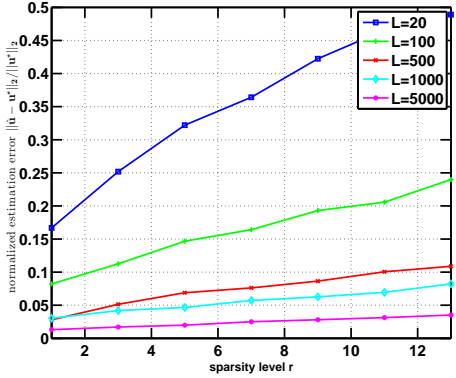}
\caption{The normalized estimation error with respect to the sparsity level $r$ for various $L$ when $n=64$ and $m=15$ for algorithm \eqref{equ:optimization}.}
\label{fig:Lcovariance}
\end{figure}

We first examine the influence of $L$ on estimating the structured covariance matrix. We fix $n=64$, and select $m=15$ entries uniformly at random from each measurement vector. The frequencies are selected uniformly from $\left[0,1\right)$, and the coefficients for each frequency are randomly drawn from $\mathcal{CN}\left(0,1\right)$. For various number of measurement vectors $L$ and sparsity level $r$, we conduct the algorithm (\ref{equ:optimization}) and record the normalized estimation error defined as $\left\Vert\hat{\boldsymbol{u}}-\boldsymbol{u}^{\star}\right\Vert_{2} /\left\Vert\boldsymbol{u}^{\star}\right\Vert_{2}$, where $\hat{\boldsymbol{u}}$ is the estimate obtained from \eqref{equ:optimization} while $\boldsymbol{u}^{\star}$ is the first column of the true covariance matrix. Each experiment is repeated 50 times, and the average normalized estimation error is calculated, which is shown in Fig. \ref{fig:Lcovariance} with respect to the sparsity level $r$ for  $L=20, 100, 500, 1000$ and $5000$. It can be seen that as $L$ increases, the average normalized estimation error decreases for a fixed sparsity level.

\begin{figure}[h]
\centering
\includegraphics[width=0.5\textwidth]{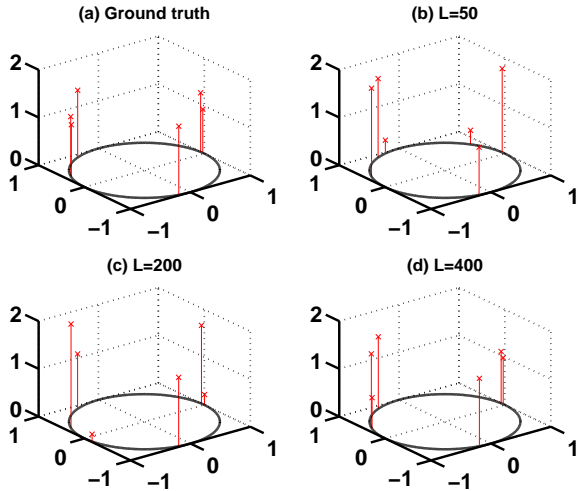}
\caption{Frequency estimation using \eqref{equ:optimization} for different $L$'s when $n=64$, $m=8$ and $r=6$. (a) Ground truth; (b) $L=50$; (c) $L=200$; and (d) $L=400$.}
\label{fig:Lfrequency}
\end{figure}

We next examine the influence of $L$ on frequency estimation using the obtained Toeplitz covariance matrix. This is done in MATLAB via applying the "rootmusic" function with the true model order (i.e. the sparsity level $r$). We fix $n=64$, and pick $m=8$ entries uniformly at random from each measurement vector. Fig.~\ref{fig:Lfrequency} (a) shows the ground truth of the set of frequencies, where the amplitude of each frequency is given as the variance in \eqref{second_order}. Fig.~\ref{fig:Lfrequency} (b)--(d) demonstrate the set of estimated frequencies when $L=50$, $200$, and $400$ respectively. As we increase $L$, the estimates of the frequencies get more accurate, especially at separating close-located frequencies. It is also worth noticing that the amplitudes of the frequencies are not as well estimated, due to the small value of $m$.

\subsection{Comparisons Between Different Approaches}\label{ssssec:covariancecompare}

The following experiment examines if more measurement vectors will lead better estimation of closely-located frequencies. Fix $n=32$ and $r=2$. In particular, we let $f_1 = 0$ and $f_2=\Delta$ which is the separation parameter. Under the same setting as Fig.~\ref{f_est_crb}, we examine the phase transition of frequency recovery for various pairs of $(\Delta, L)$. For each Monte Carlo simulation, it is considered successful if $\sum_{k=1}^{r}\left(\hat{f}_{k}-f_{k}\right)^{2}/r\le 10^{-5}$, where $\hat{f}_{k}$ is the estimate of $f_{k}$. We implement the two proposed algorithms with full observations and half randomly-selected observations respectively. 

\begin{figure}[ht]
\centering
\begin{tabular}{cc}
\hspace{-0.2in}\includegraphics[width=0.27\textwidth]{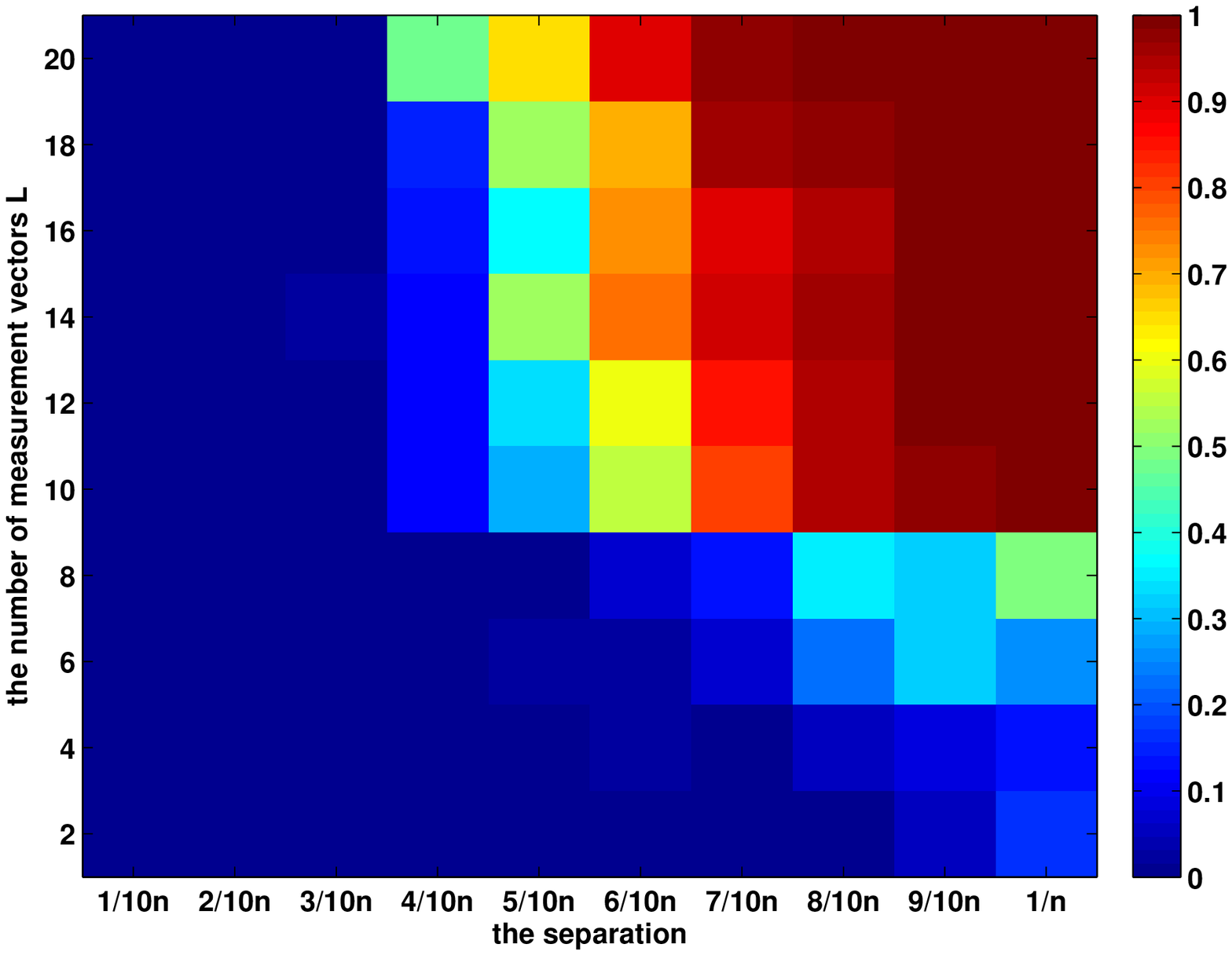} &
\hspace{-0.3in}\includegraphics[width=0.27\textwidth]{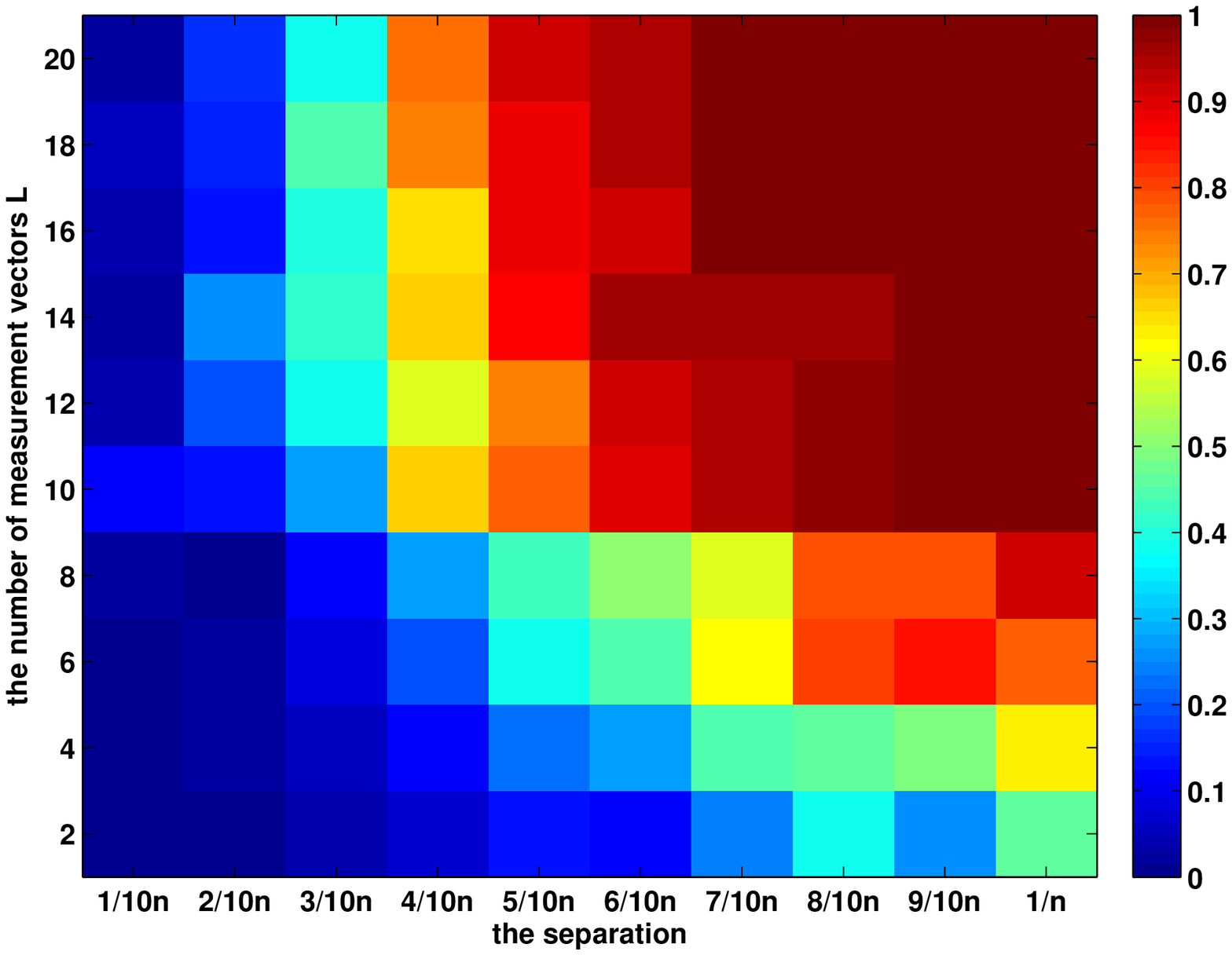}\\
\hspace{-0.2in}{\scriptsize \tabincell{c}{(a) Atomic norm minimization \\ with full observations}}  & 
\hspace{-0.3in}{\scriptsize \tabincell{c}{ (b) Structured covariance estimation \\ with full observations}}  \\
\hspace{-0.2in}\includegraphics[width=0.27\textwidth]{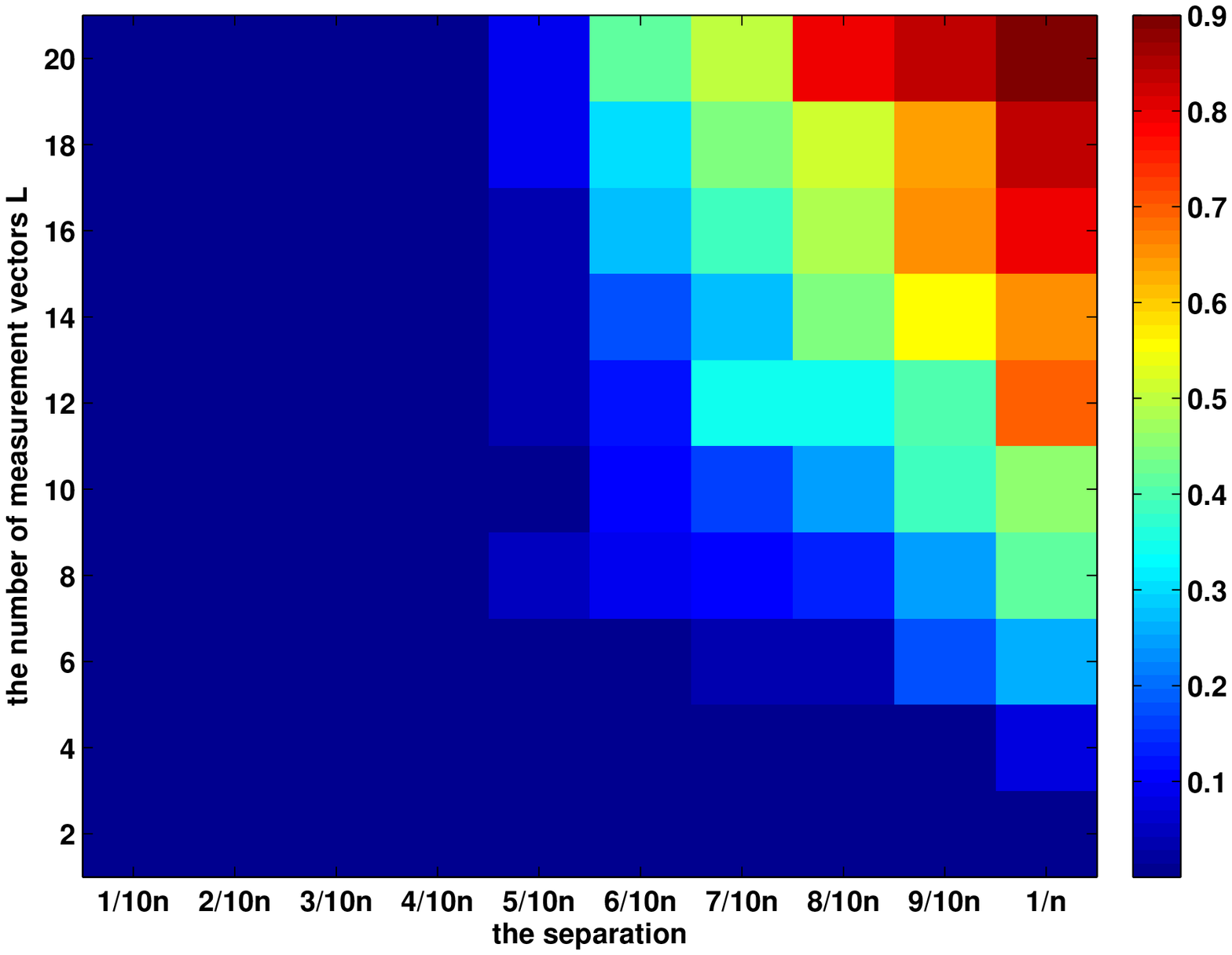} &
\hspace{-0.3in}\includegraphics[width=0.27\textwidth]{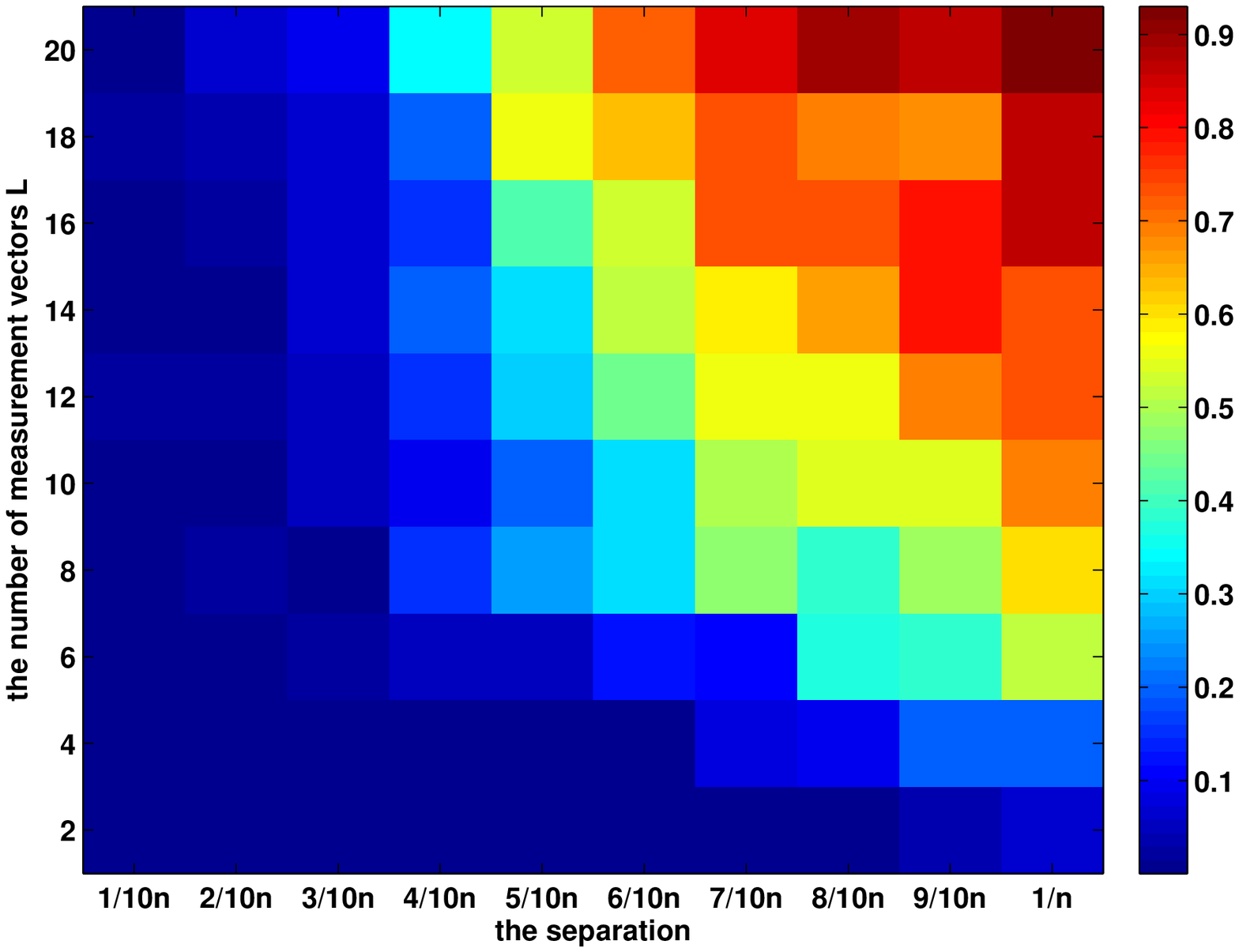}\\
\hspace{-0.2in} {\scriptsize \tabincell{c}{(c) Atomic norm minimization \\ with half randomly-selected observations}  } &
\hspace{-0.3in} {\scriptsize \tabincell{c}{(d) Structured covariance estimation \\ with half randomly-selected observations }  }
\end{tabular}
\caption{Phase transitions of the proposed algorithms for frequency estimation with respect to the number of measurement vectors $L$ and the separation parameter when $n=32$, $r=2$ and $\sigma=0.3$.}
\label{fig:phase_trans_L_separation_noisy}
\end{figure}

Fig.~\ref{fig:phase_trans_L_separation_noisy} shows the successful rate of frequency estimation for atomic norm minimization in (a) and (c), and for structured covariance estimation in (b) and (d). Indeed, the success rate increases as one increases $L$ for a fixed separation parameter. Alternatively, to achieve the same success rate, a smaller separation is possible with a larger $L$. Furthermore, the performance also increases as more samples per measurement vector is available. The structure covariance estimation approach achieves better phase transition compared to the atomic norm minimization approach.



\begin{figure*}[ht]
\centering
\begin{tabular}{ccccc}
\includegraphics[width=0.2\textwidth]{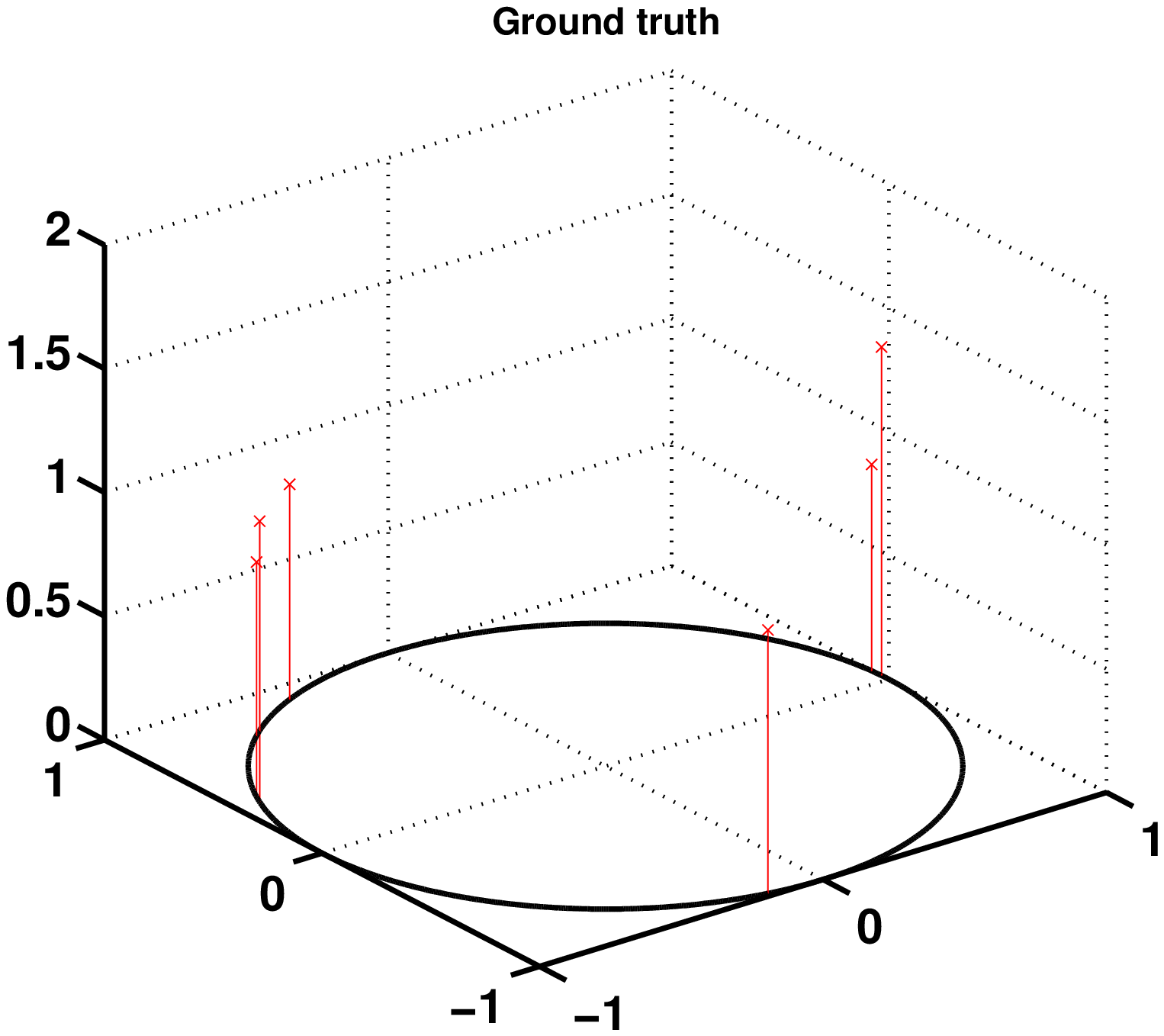} &
\hspace{-0.2in}\includegraphics[width=0.2\textwidth]{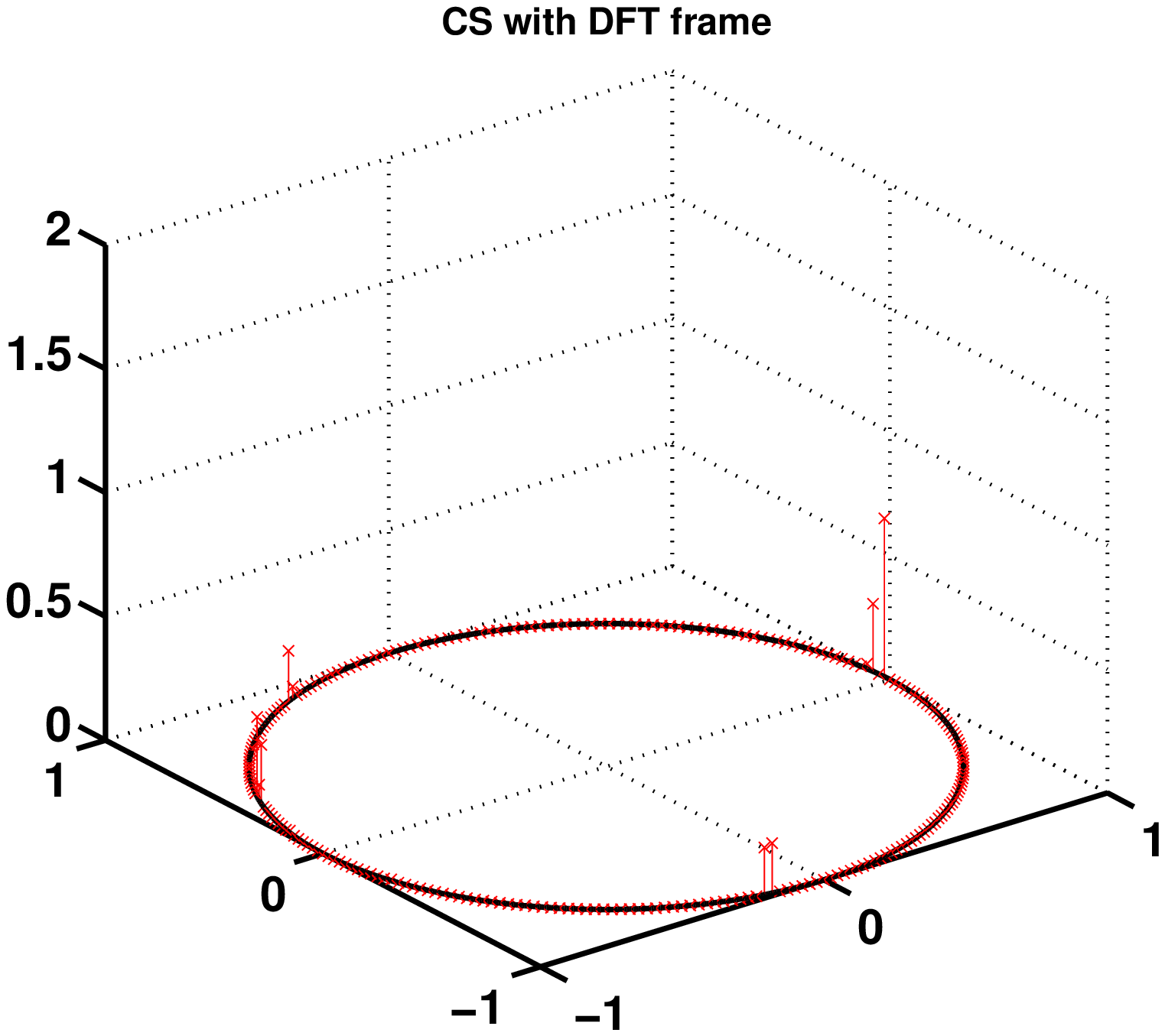}&
\hspace{-0.2in}\includegraphics[width=0.2\textwidth]{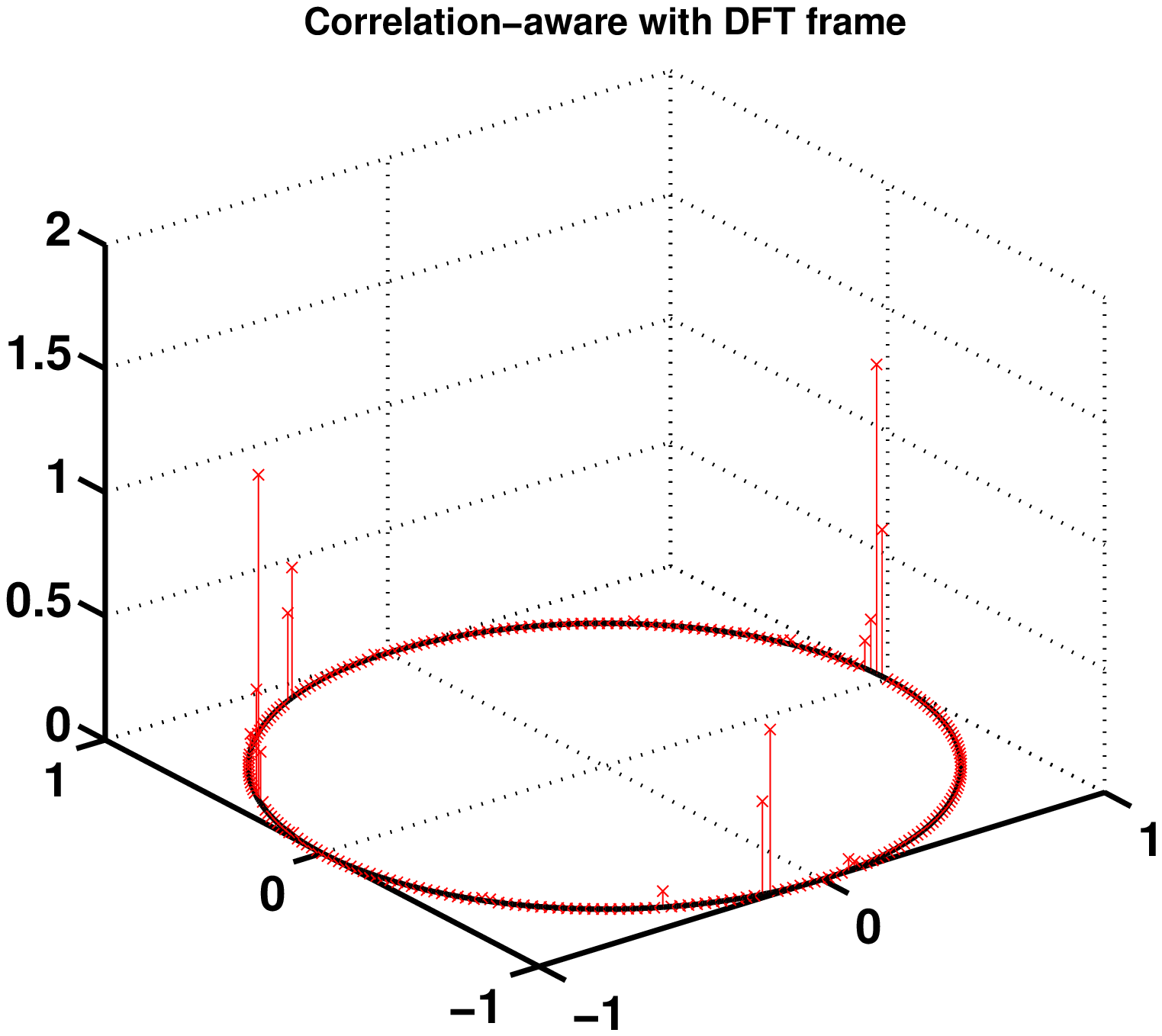} & \hspace{-0.2in} \includegraphics[width=0.2\textwidth]{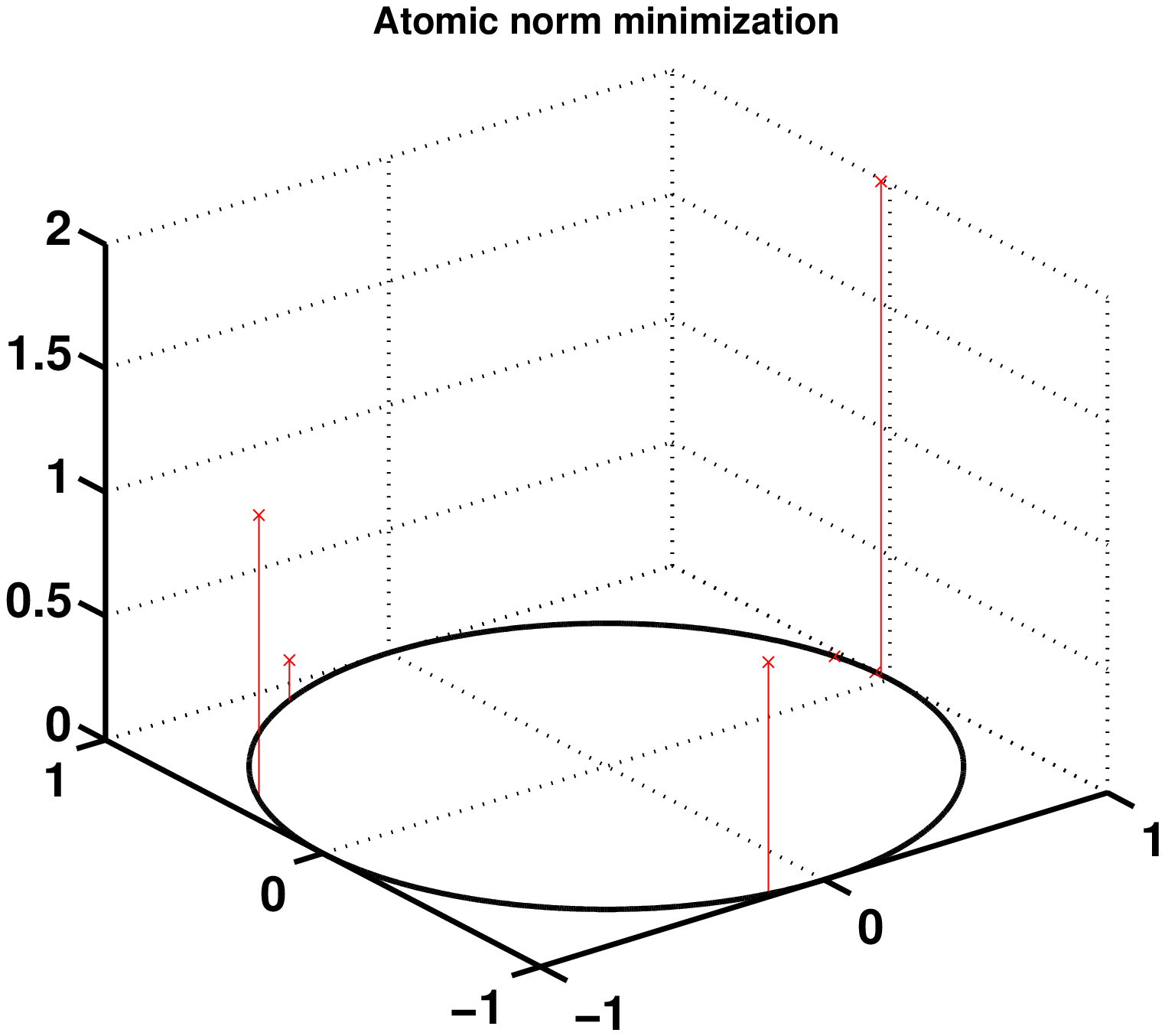}&\hspace{-0.2in}
\includegraphics[width=0.2\textwidth]{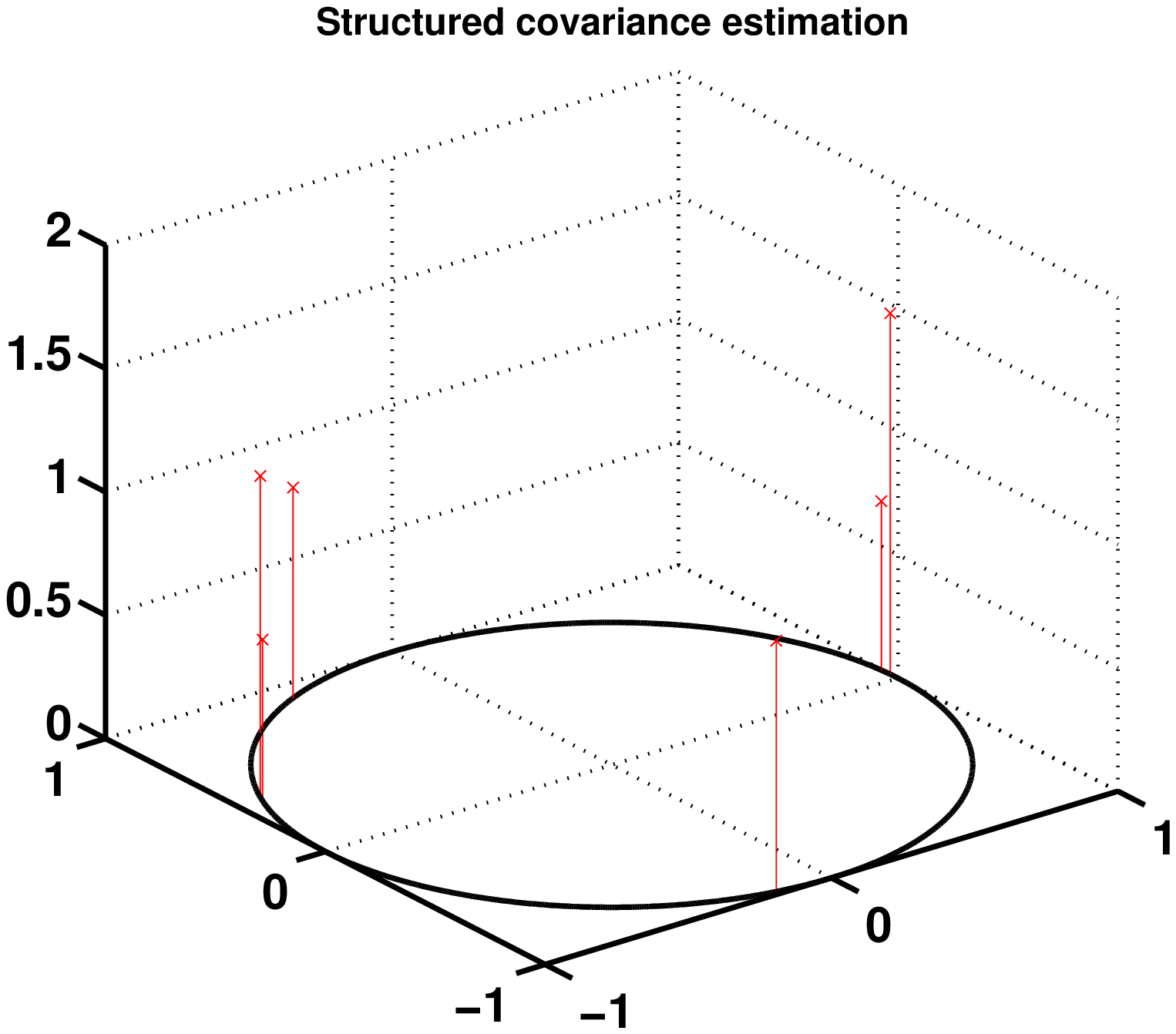} \\
\includegraphics[width=0.2\textwidth]{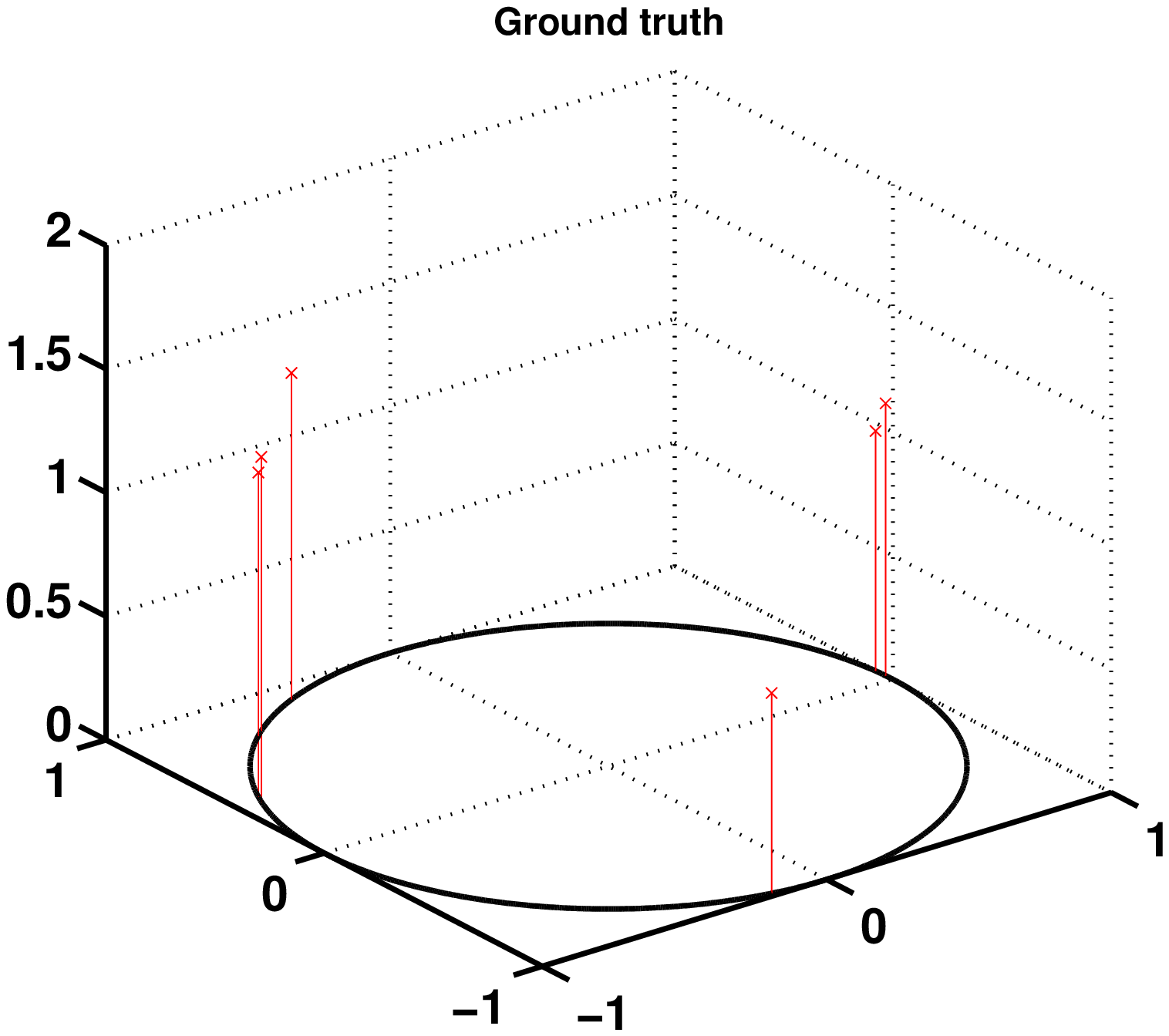} &
\hspace{-0.2in}\includegraphics[width=0.2\textwidth]{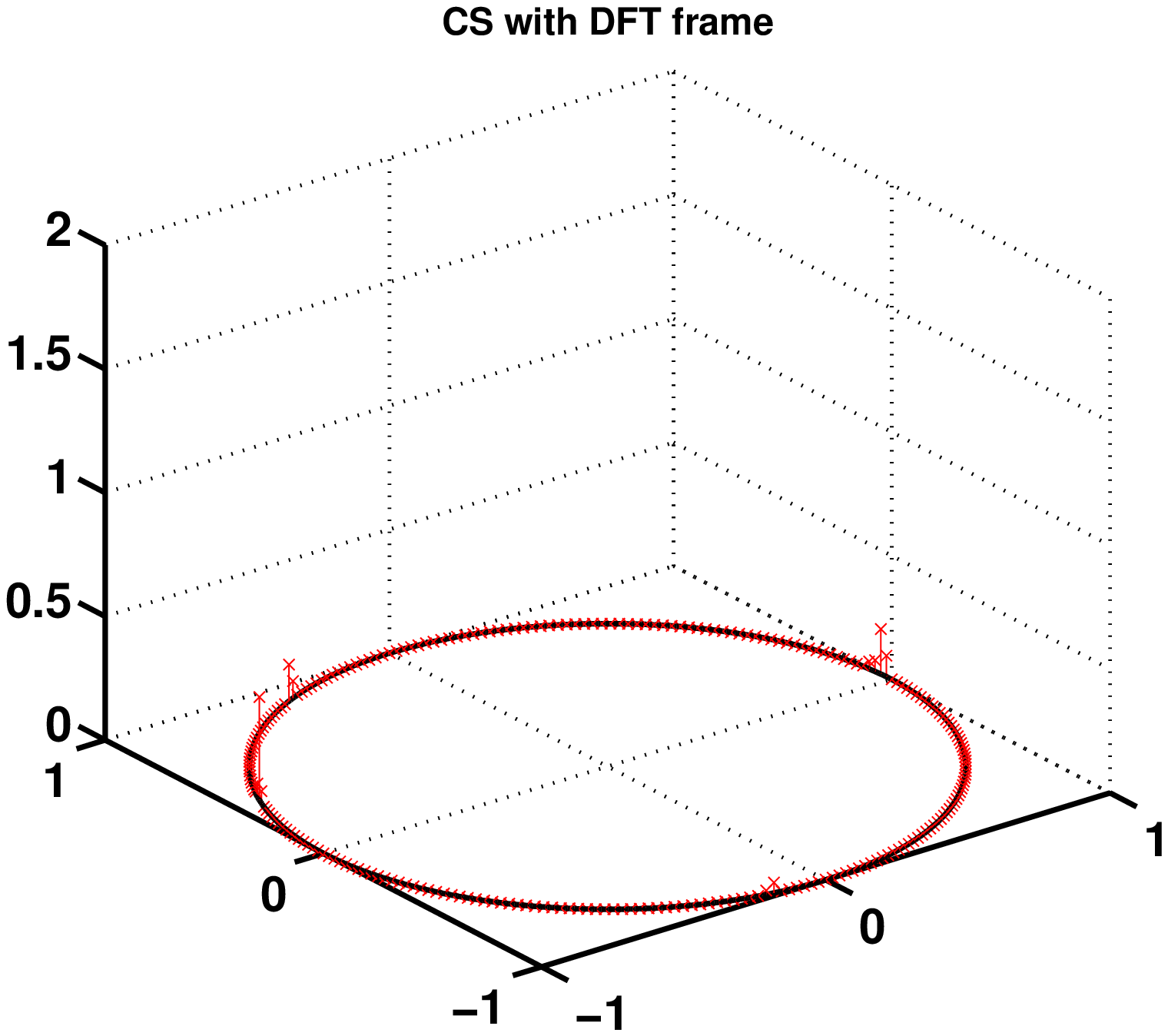}&
\hspace{-0.2in}\includegraphics[width=0.2\textwidth]{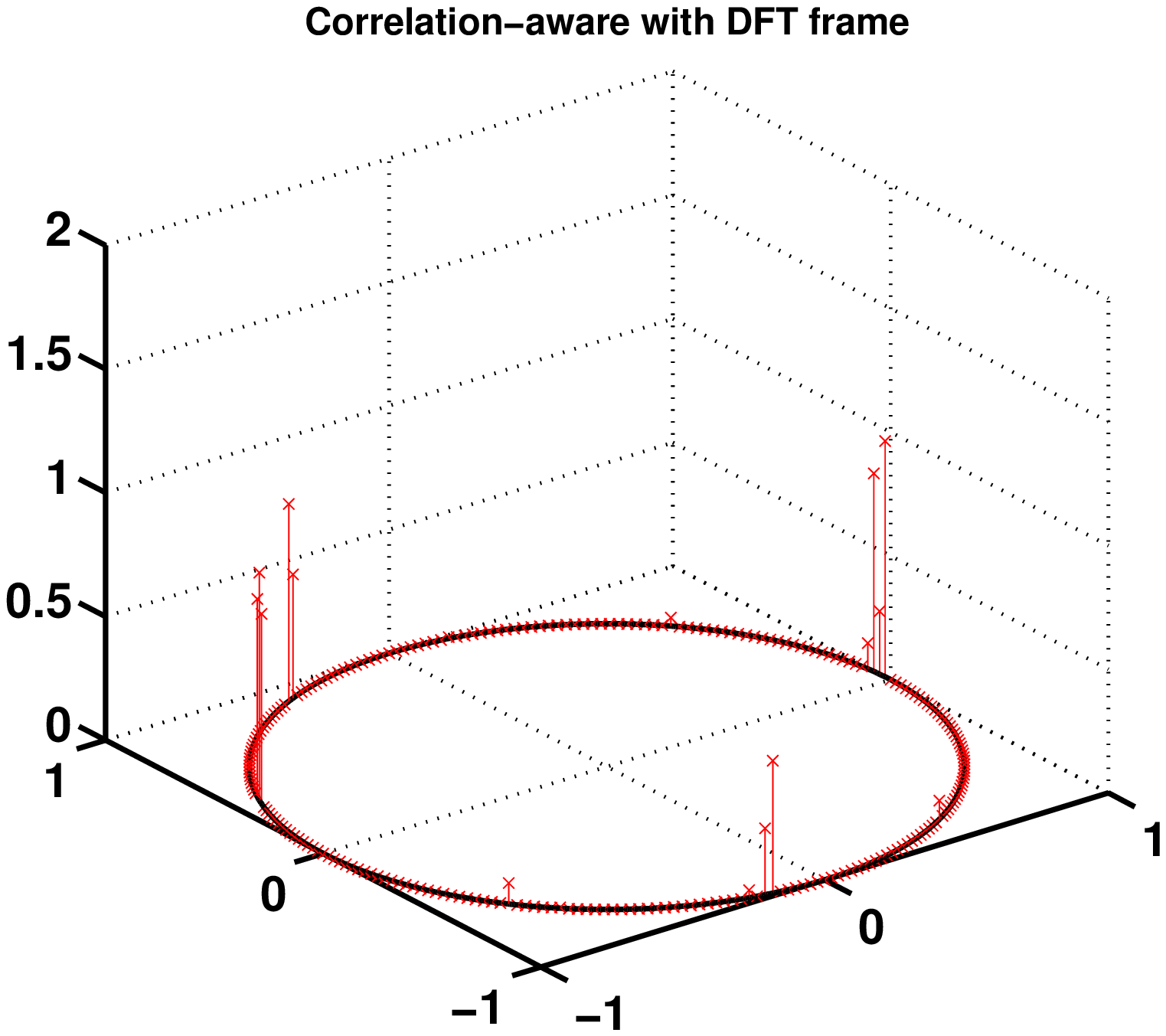} & \hspace{-0.2in} \includegraphics[width=0.2\textwidth]{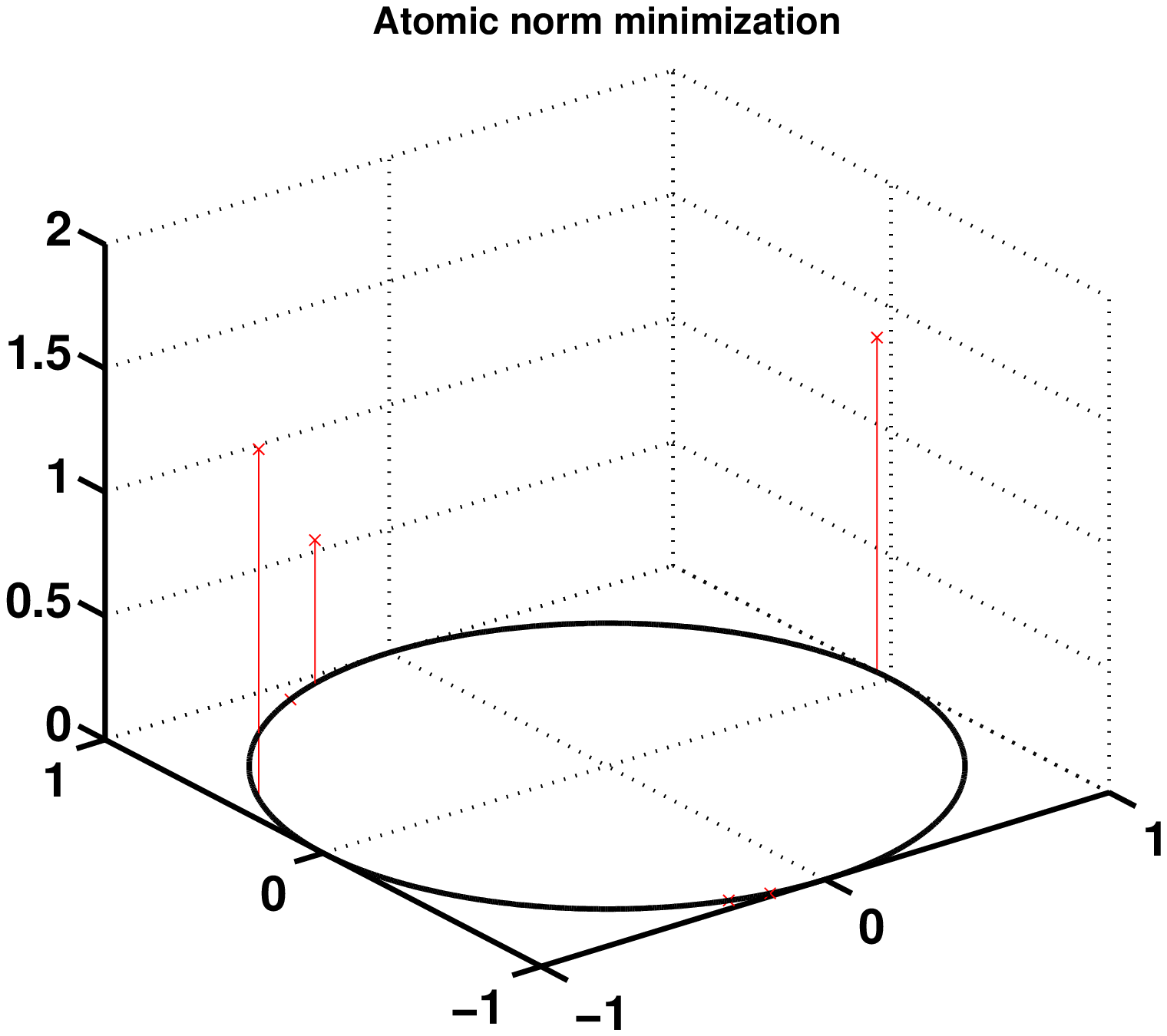}&\hspace{-0.2in}
\includegraphics[width=0.2\textwidth]{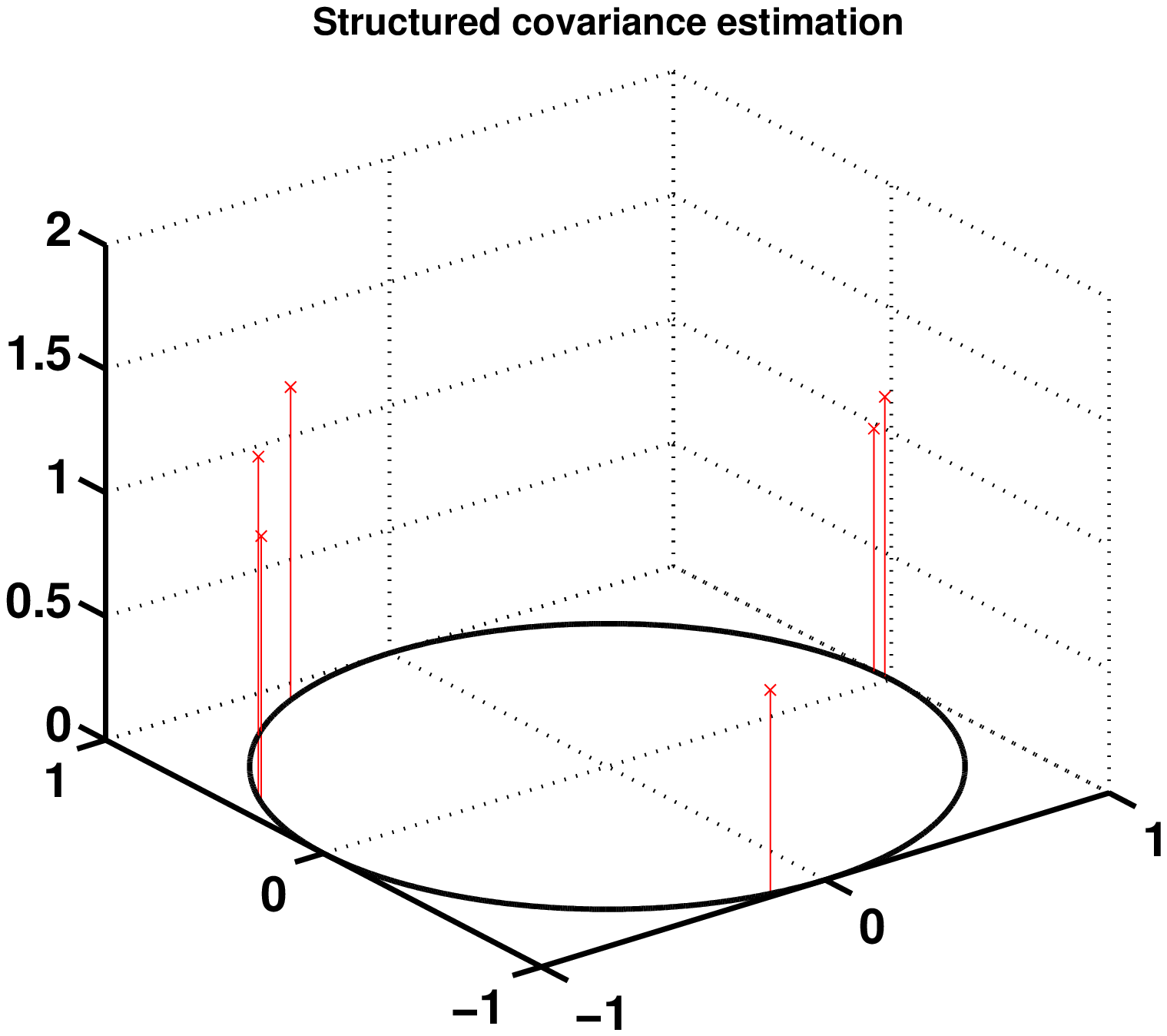} \\
{\scriptsize (a)}  & \hspace{-0.2in}{\scriptsize (b) }  & \hspace{-0.2in} {\scriptsize (c)}  &   \hspace{-0.2in}{\scriptsize (d)}   &  \hspace{-0.2in} {\scriptsize (e)}  
\end{tabular}
\caption{Frequency estimation with noiseless measurements using different algorithms when $n=64$ and $L=400$. First row: $m=8$, $r=6$; Second row: $m=5$, $r=6$. (a) Ground truth; (b) CS with the DFT frame; (c) Correlation-aware with the DFT frame; (d) Atomic norm minimization; (e) Structured covariance estimation.}
\label{fig:frequencycompare}
\end{figure*}
\begin{figure*}[ht]
\centering
\begin{tabular}{ccccc}
\includegraphics[width=0.2\textwidth]{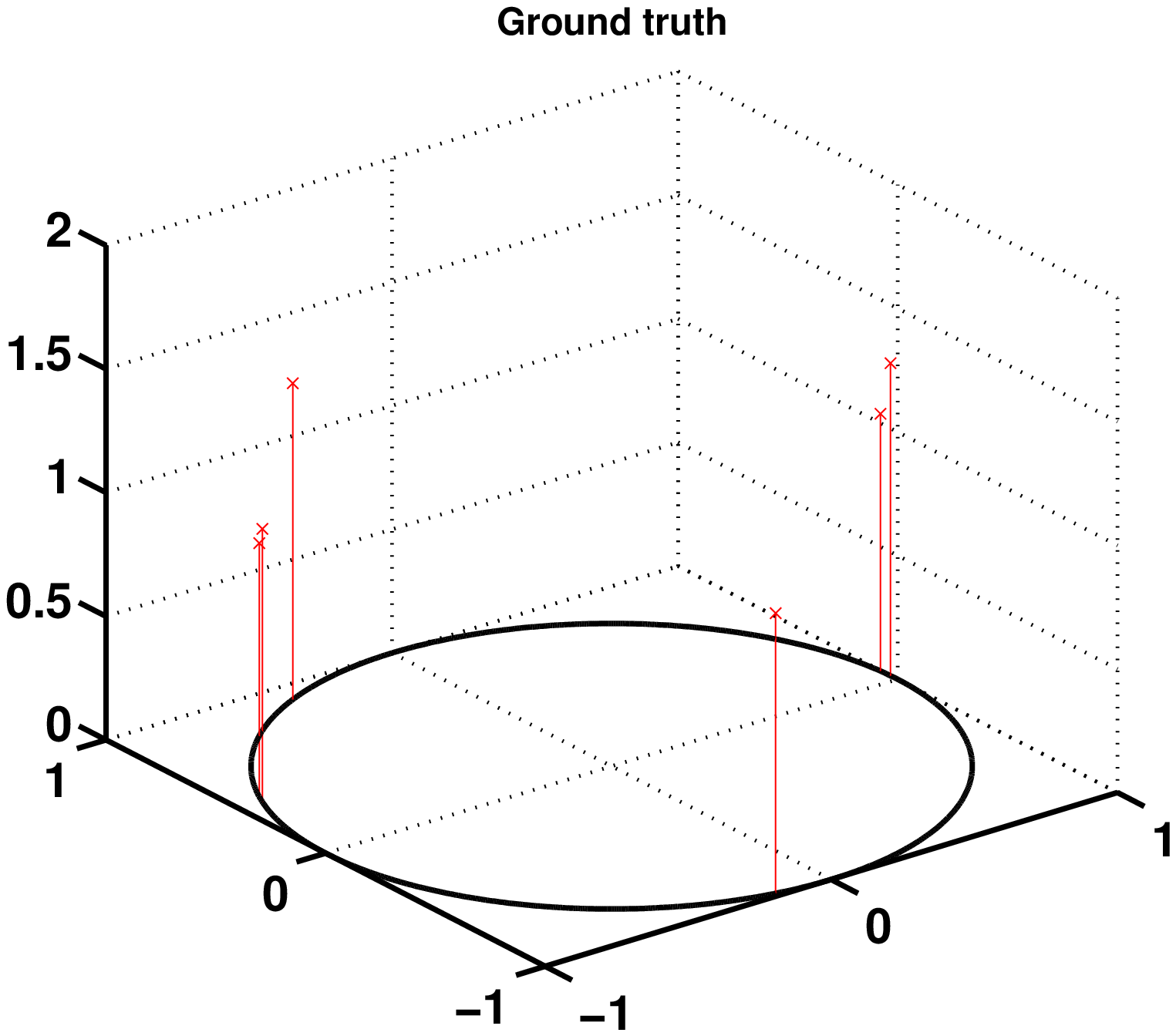} &
\hspace{-0.2in}\includegraphics[width=0.2\textwidth]{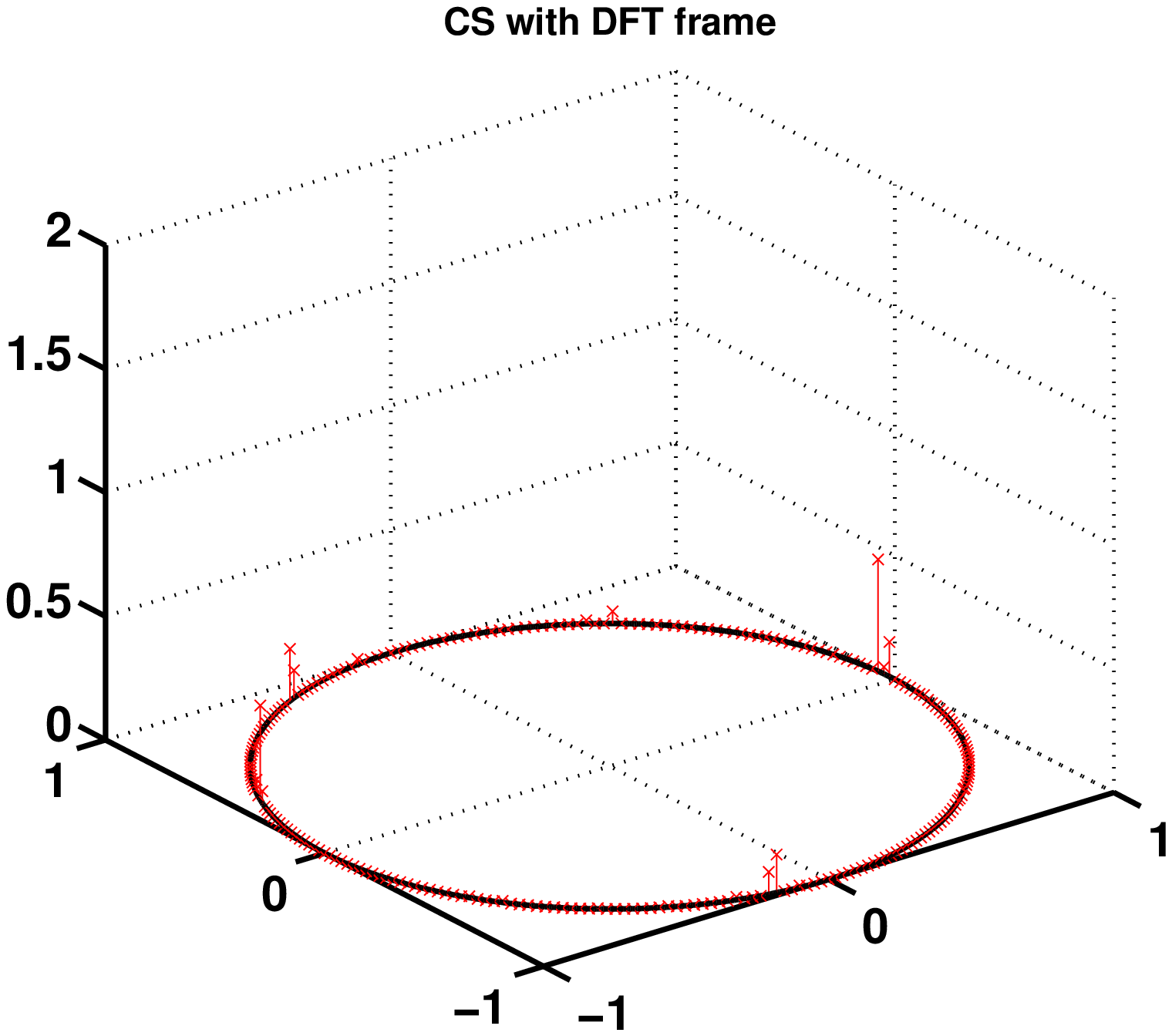}&
\hspace{-0.2in}\includegraphics[width=0.2\textwidth]{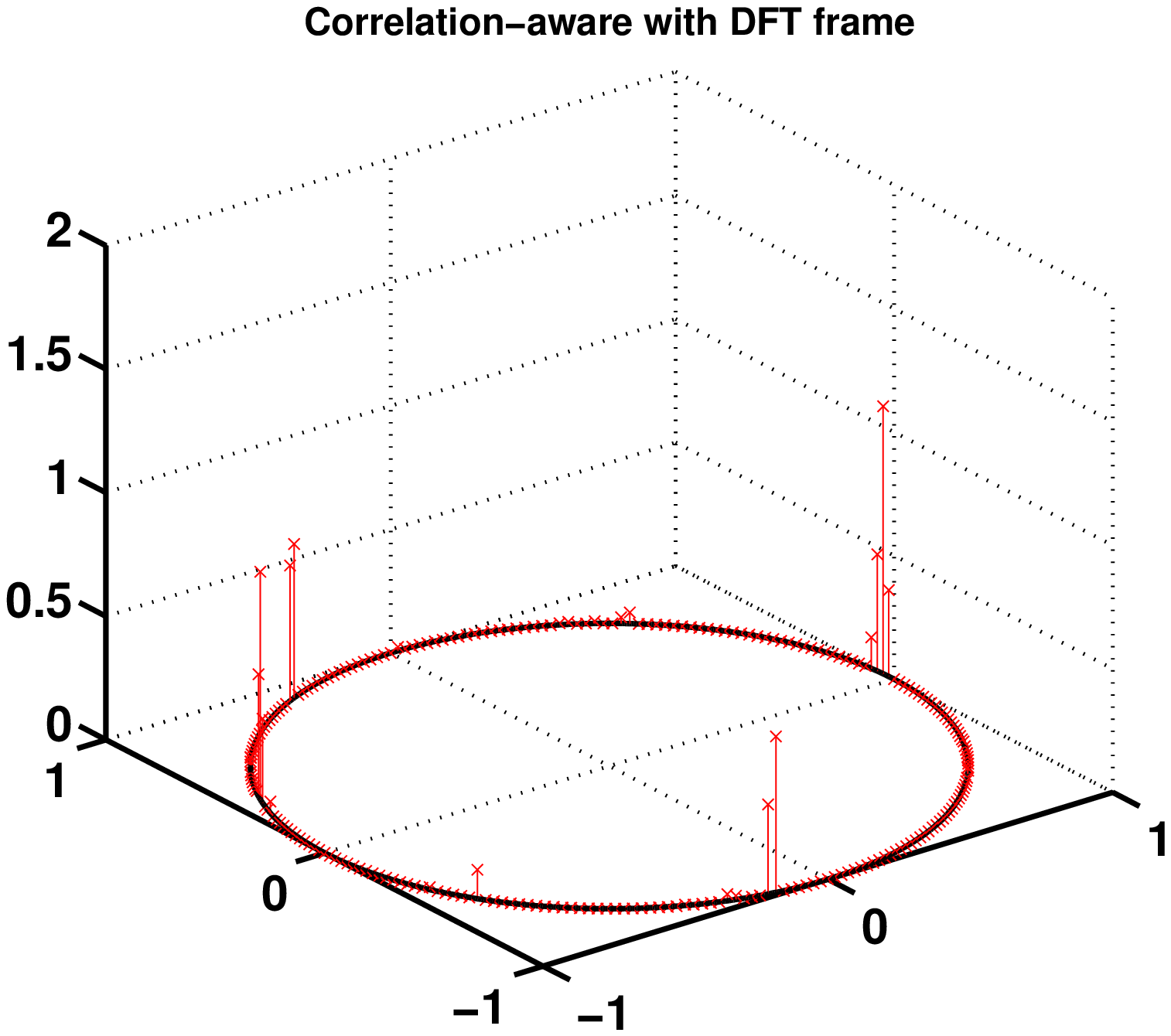} & \hspace{-0.2in} \includegraphics[width=0.2\textwidth]{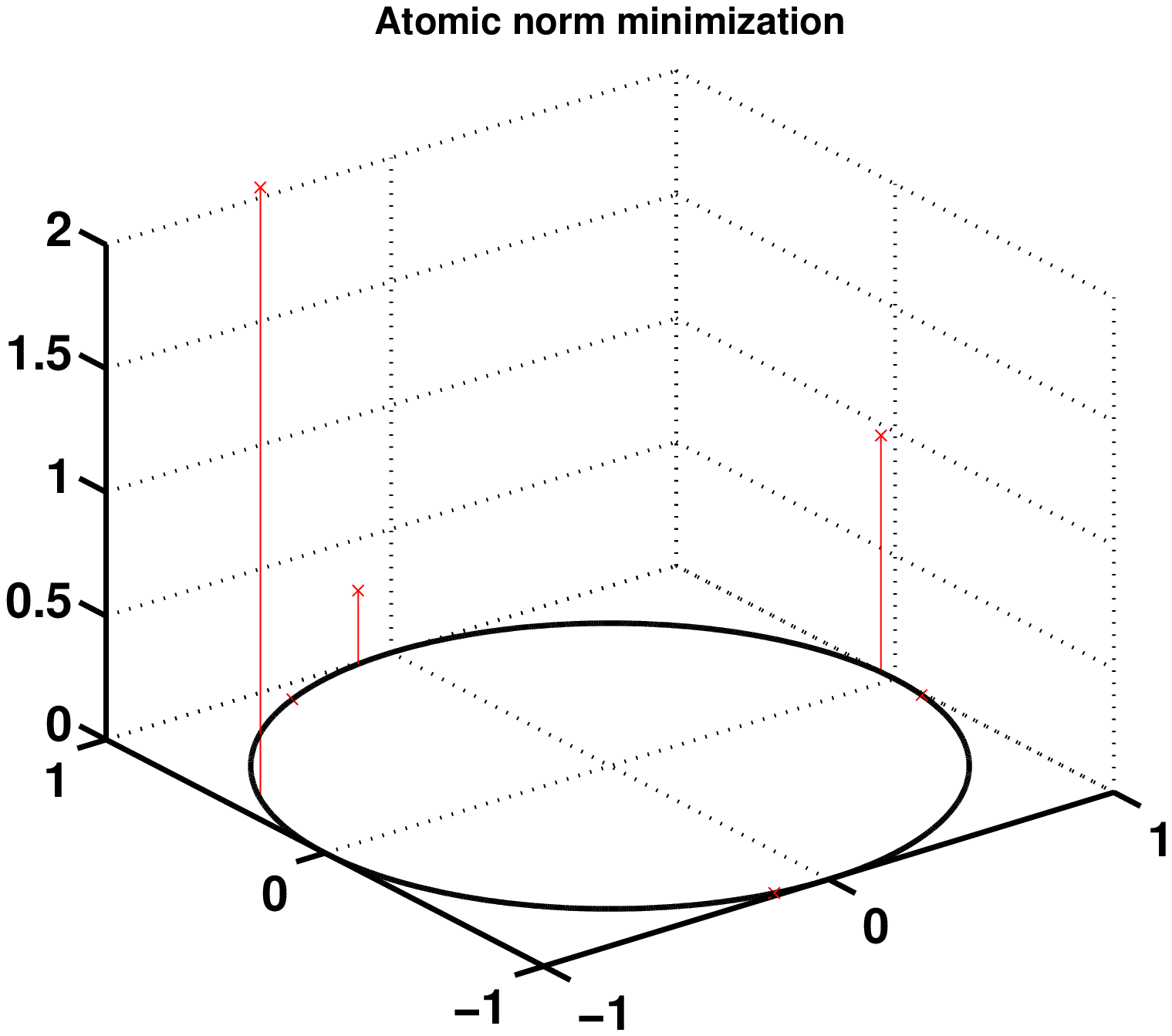}&\hspace{-0.2in}
\includegraphics[width=0.2\textwidth]{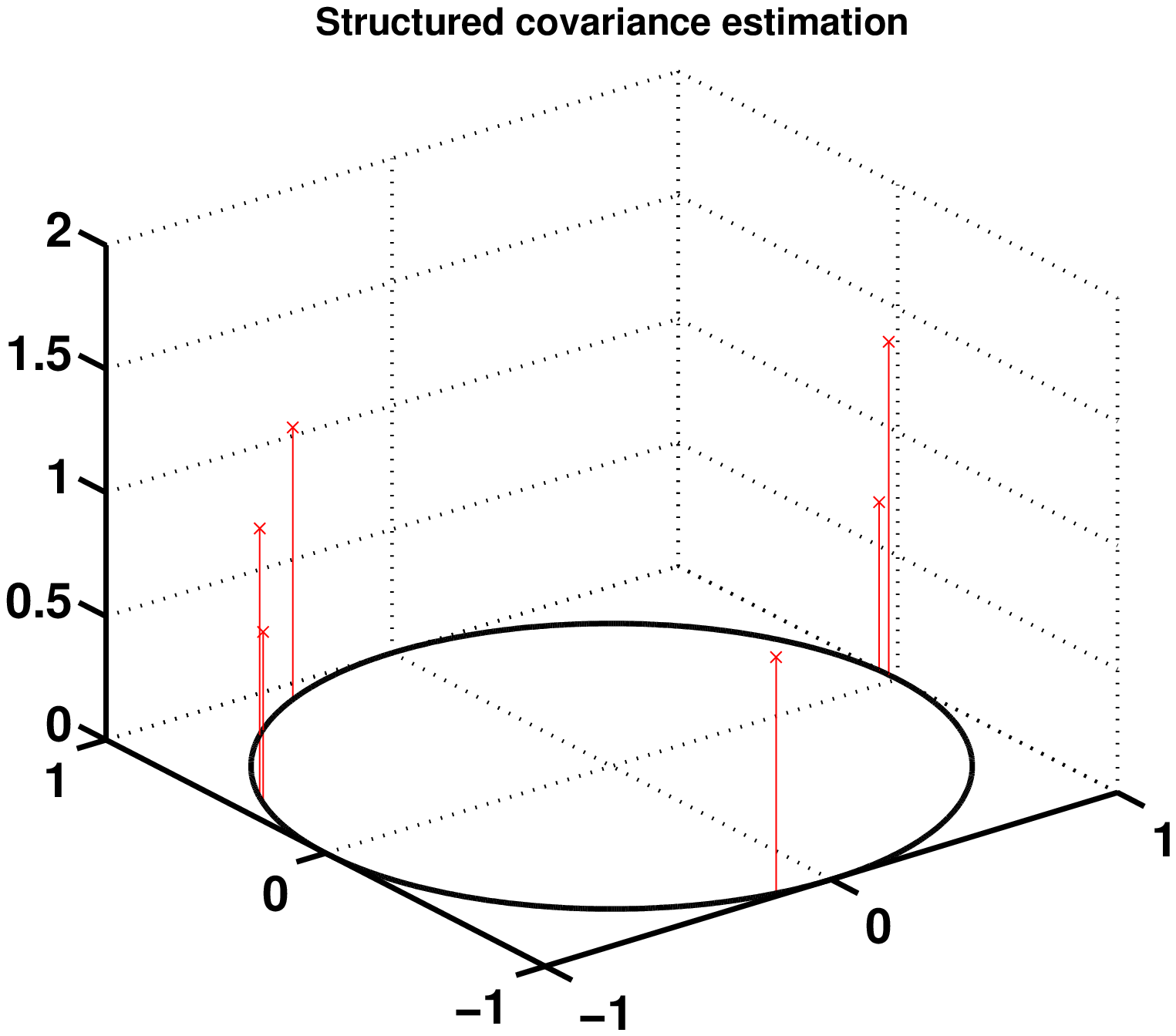} \\
\includegraphics[width=0.2\textwidth]{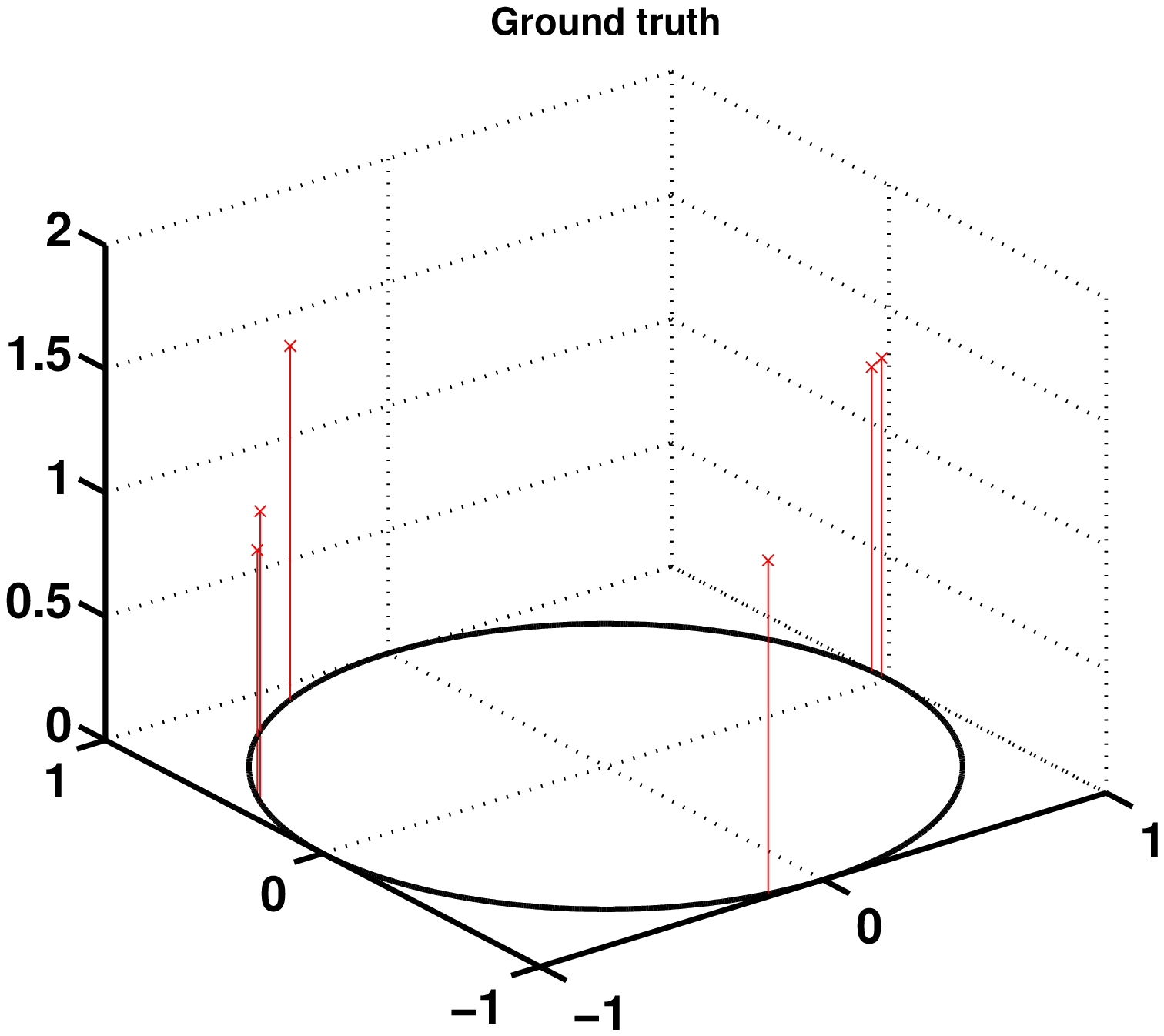} &
\hspace{-0.2in}\includegraphics[width=0.2\textwidth]{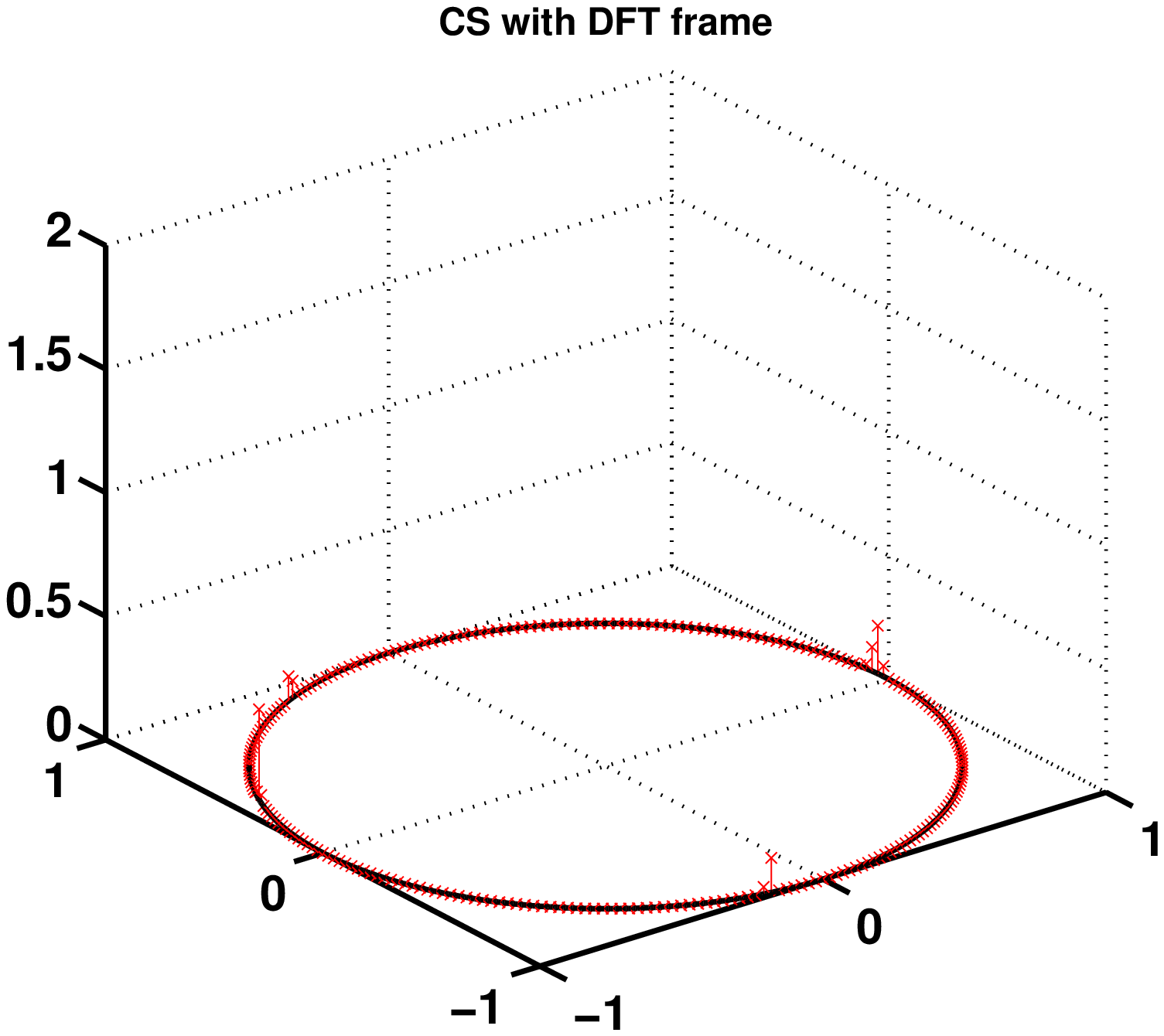}&
\hspace{-0.2in}\includegraphics[width=0.2\textwidth]{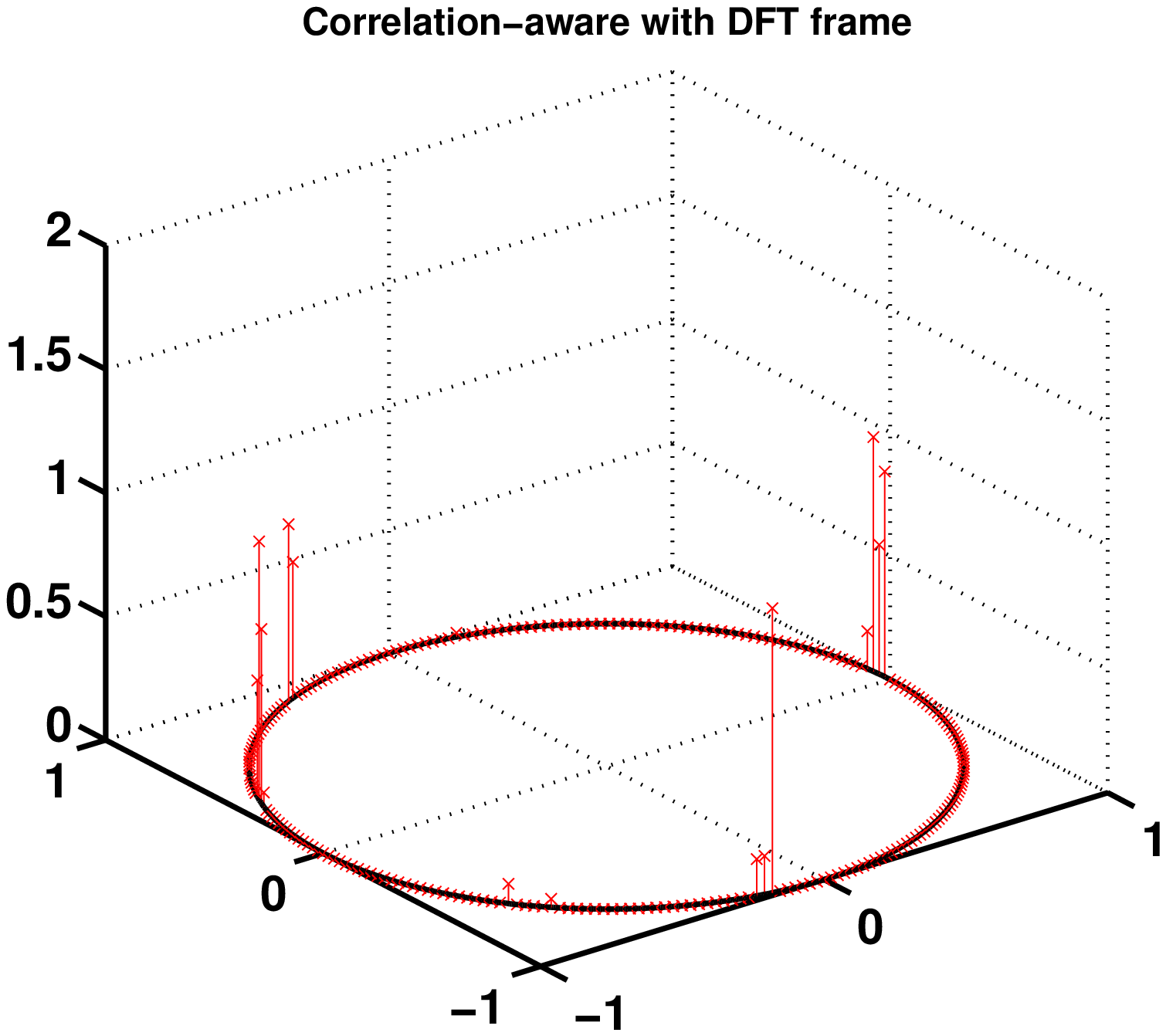} & \hspace{-0.2in} \includegraphics[width=0.2\textwidth]{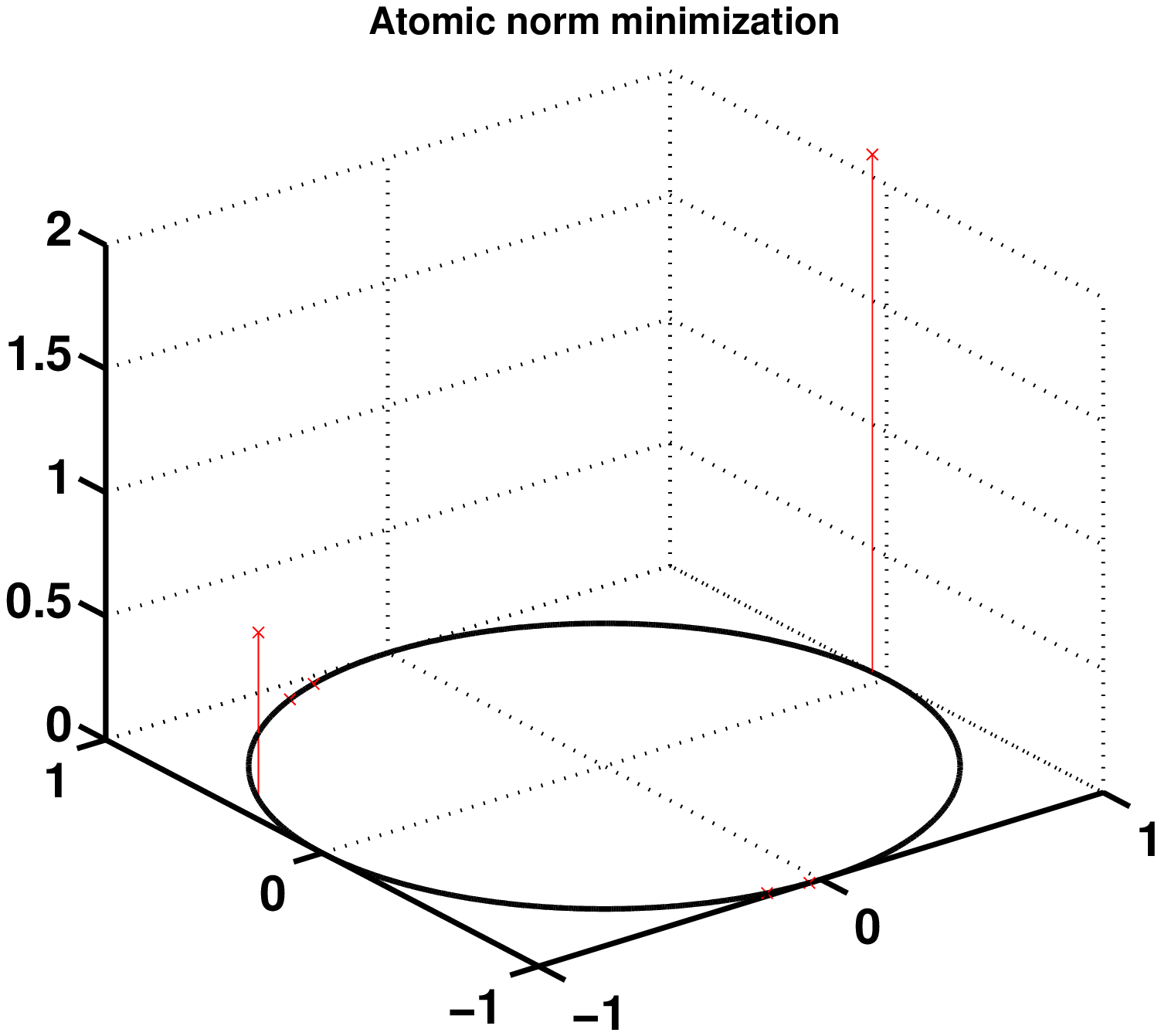}&\hspace{-0.2in}
\includegraphics[width=0.2\textwidth]{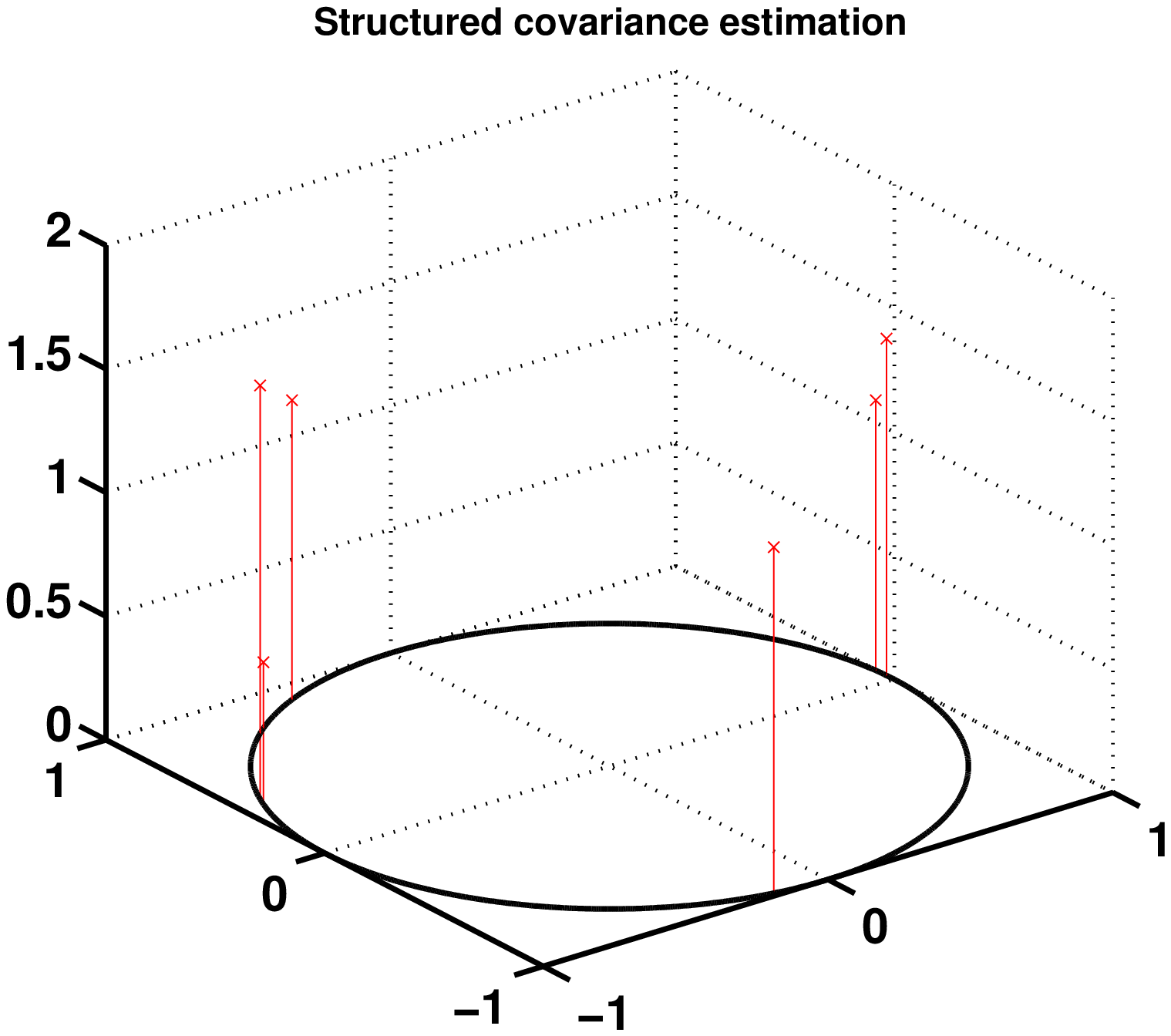} \\
{\scriptsize (a)}    & \hspace{-0.2in}{\scriptsize (b)}   & \hspace{-0.2in} {\scriptsize (c)}  &   \hspace{-0.2in}{\scriptsize (d)}  &  \hspace{-0.2in} {\scriptsize (e)}  
\end{tabular}
\caption{Frequency estimation with measurements corrupted by noise $\mathcal{CN}\left(0,0.2^{2}\right)$ using different algorithms when $n=64$ and $L=400$. First row: $m=8$, $r=6$; Second row: $m=5$, $r=6$. (a) Ground truth; (b) CS with the DFT frame; (c) Correlation-aware with the DFT frame; (d) Atomic norm minimization; (e) Structured covariance estimation.}
\label{fig:frequencycompare_noise}
\end{figure*}

We first compare qualitatively the performance of frequency estimation using different algorithms, including CS using group sparsity with a DFT frame \cite{tropp2006algorithms}, the correlation-aware approach \cite{pal2012application}, atomic norm minimization \eqref{primal}, and structured covariance estimation \eqref{equ:optimization}. For CS and correlation-aware method, we assume a DFT frame with an oversampling factor $4$. For the correlation-aware method, we empirically set its regularization parameter as $h=2\times 10^{-4}/\left(\log{L}\cdot\log{m}\right)^{2}$ which gives good performance \cite{pal2012application}.

Let $n=64$, $L=400$ and $r=6$. We generate a spectrally sparse ground truth scene in Fig.~\ref{fig:frequencycompare} (a) in the same way as Fig.~\ref{fig:Lfrequency} (a). Fig.~\ref{fig:frequencycompare} (b)--(e) respectively show the estimated frequencies on a unit circle for different methods, with $m=8$ and $m=5$ at $\Omega=\{0,32,39,47,57\}$ respectively in the first row and the second row. The structured covariance estimation algorithm works well to locate all the frequencies accurately in both cases. Due to the off-the-grid mismatch, CS and correlation-aware techniques predict frequencies on the lattice of the DFT frame, and result in a larger number of estimated frequencies. On the other hand, atomic norm minimization fails to distinguish the two close frequencies and misses one frequency due to insufficient number of measurements per vector. 
We then repeat the experiment of Fig.~\ref{fig:frequencycompare} where the signals are corrupted by AWGN $\mathcal{CN}\left(0,\sigma^{2}\right)$, where $\sigma=0.2$. Fig.~\ref{fig:frequencycompare_noise} shows the performance of each method in a unit circle. Notice that the structured covariance estimation algorithm can still work well to locate all the frequencies accurately, despite there is certain inaccuracy on the corresponding amplitude estimation.

\begin{figure}[htp]
\begin{center}
\begin{tabular}{cc}
\hspace{-0.3in}\includegraphics[height= 1.7in, width=0.28\textwidth]{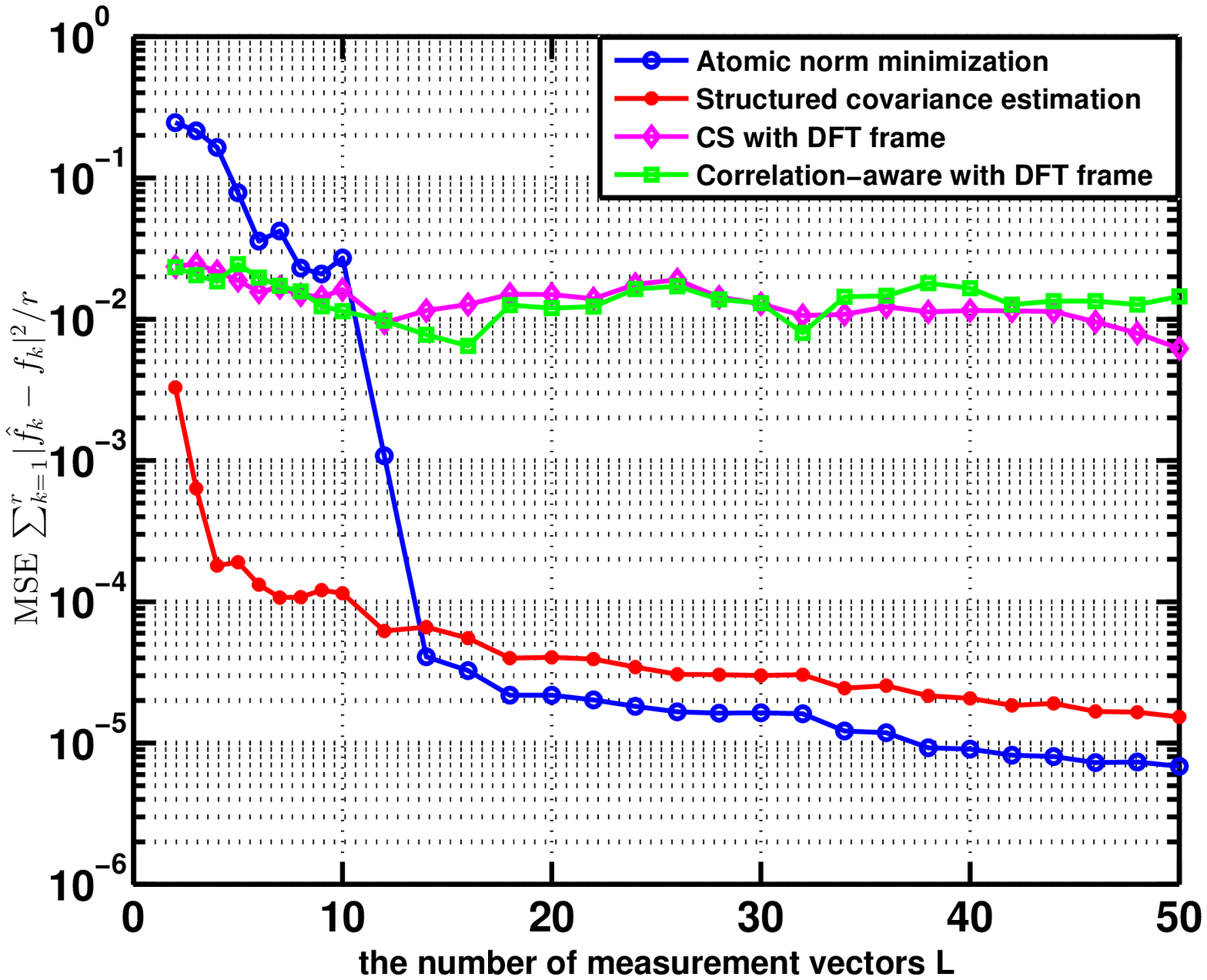} &\hspace{-0.34in} \includegraphics[height= 1.7in, width=0.28\textwidth]{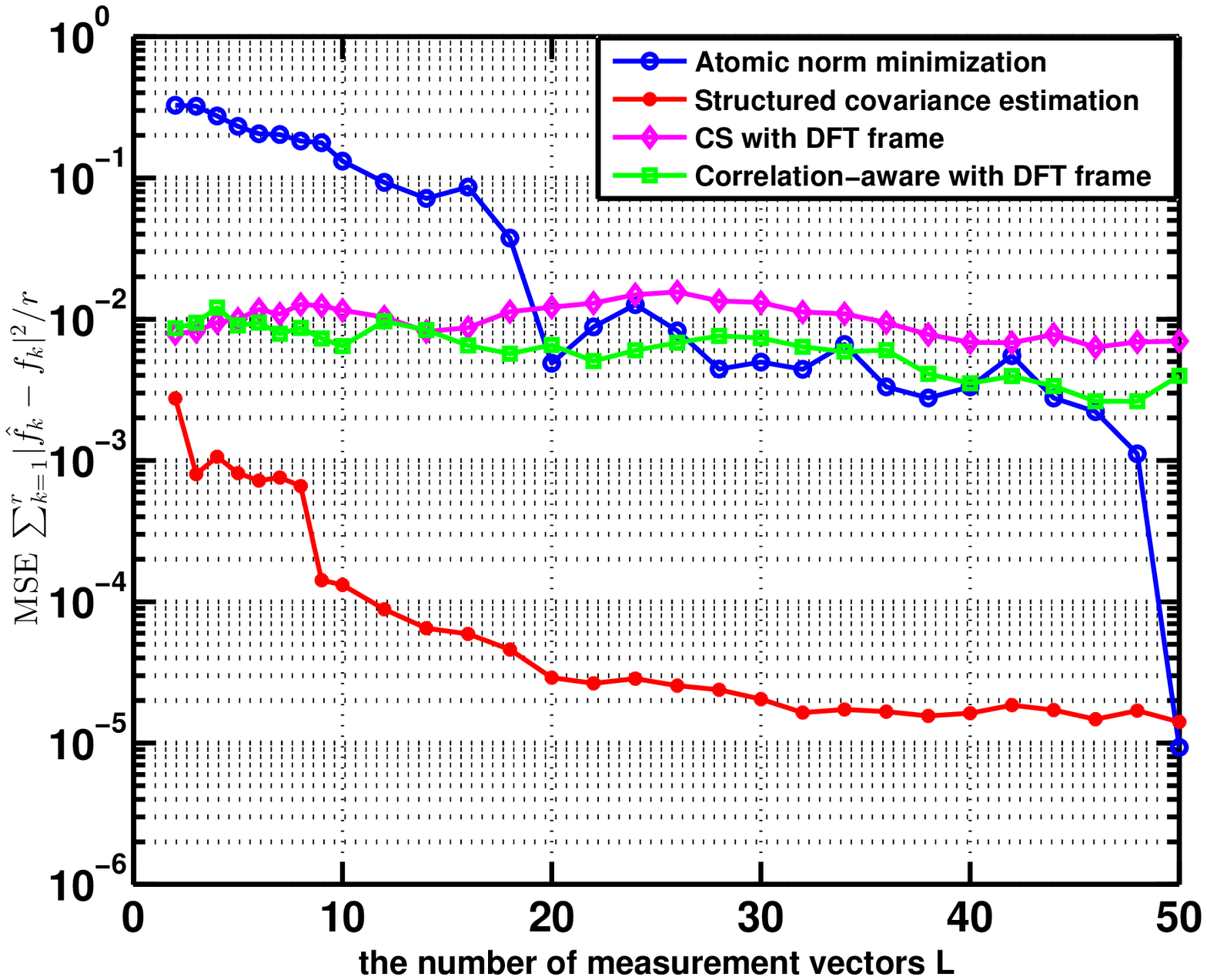}  \\
{\scriptsize \hspace{-0.3in}(a) Separation condition is satisfied} &
{\scriptsize \hspace{-0.3in}(b) Separation condition is not satisfied}
\end{tabular}
\end{center}
\caption{The comparisons of average frequency estimation MSE's with respect to $L$ when $n=12$, $r=8$ and $\sigma=0.3$.} \label{fig:fre_est_mse_compare}
\end{figure}

We next compare the average performance of frequency estimation between different algorithms. Let $n=12$ and $r=8$. We fix a set of frequencies that satisfies the separation condition $\Delta\ge 1/n$. The coefficient matrix $\boldsymbol{C}$ is generated with i.i.d. $\mathcal{CN}\left(0,1\right)$ entries, and the noise matrix $\boldsymbol{N}$ is generated with i.i.d. $\mathcal{CN}\left(0, 0.3^{2}\right)$ entries. Fig.~\ref{fig:fre_est_mse_compare} (a) shows the average MSE, calculated as $\sum_{k=1}^{r}\left(\hat{f}_{k}-f_{k}\right)^{2}/r$,  where $\hat{f}_{k}$ is the estimate of $f_{k}$ for different algorithms over 200 Monte Carlo trials. It can be seen that the structured covariance estimation algorithm achieves superior performance when $L$ is small, while the atomic norm minimization algorithm dramatically improves its performance as soon as $L$ is large enough, and both are much better than the grid-based approaches. We then on purposely move two pairs of frequencies to violate the separation condition, and rerun the same simulation. Fig.~\ref{fig:fre_est_mse_compare} (b) shows the average MSE under this setting, where similar behaviors are observed. However, the atomic norm minimization algorithm requires more measurement vectors in order to approach the performance of structured covariance estimation.

\section{Conclusion and Future Work}\label{sec:conclusion}
 
In this paper, we study the problem of line spectrum estimation and denoising of multiple spectrally-sparse signals from their possibly partial and noisy observations, where all the signals are composed of a common set of continuous-valued frequencies. Two approaches are developed and solved efficiently via semidefinite programming. The first algorithm aims to recover the signal ensemble based on atomic norm minimization, which has a higher computational cost when the number of measurement vectors is high. The second algorithm aims to recover the structured covariance matrix from the partially observed sample covariance matrix. The set of frequencies can be recovered either via characterization of the dual polynomial, or using directly traditional methods such as MUSIC. Theoretical performance guarantees are derived for both approaches under different scenarios. The effectiveness of the proposed methods are further demonstrated through numerical examples with comparisons against existing approaches.

We outline a few future research directions, such as deriving theoretical performance guarantees for atomic norm minimization from partially observed noisy observations. Moreover, it is desirable to study the fundamental trade-offs between the separation condition, the number of measurement vectors, and the noise level using convex optimization based techniques.


\appendix
\subsection{Proof of Theorem~\ref{atomic-sdp}}\label{proof_atomic_sdp}
\begin{proof} Denote the value of the right hand side as $\|\bX\|_\cT$. Suppose the atomic decomposition of $\bX$ is given as $\bX =\sum_{k=1}^r c_k \ba(f_k)\bb_k^*$. By the Vandermonde decomposition lemma \cite{Caratheodory}, there exists a vector $\bu$ such that
$ \toep(\bu) = \sum_{k=1}^r c_k \ba(f_k) \ba(f_k)^*$. It is obvious that
\begin{align*} 
& \begin{bmatrix}
\toep(\bu) & \bX \\
\bX^* & \sum_{k=1}^r c_k \bb_k \bb_k^* \end{bmatrix} \\
& =  \begin{bmatrix}
\sum_{k=1}^r c_k \ba(f_k) \ba(f_k)^* & \sum_{k=1}^r c_k\ba(f_k) \bb_k^* \\
 \sum_{k=1}^r  c_k \bb_k\ba(f_k)^* & \sum_{k=1}^r c_k \bb_k \bb_k^* \end{bmatrix} \\
 & = \sum_{k=1}^r  c_k  \begin{bmatrix}
  \ba(f_k)  \\
  \bb_k    \end{bmatrix} \begin{bmatrix}
  \ba(f_k)^* &  \bb_k^*    \end{bmatrix} \succeq 0,
 \end{align*}
 and $ \frac{1}{2}\trace(\toep(\bu)) + \frac{1}{2}\trace(\bW) = \sum_{k=1}^r c_k = \| \bX\|_\cA$,
therefore $\|\bX\|_{\cT}\leq \|\bX\|_\cA$. On the other hand, suppose that for any $\bu$ and $\bW$ that satisfy
$$ \begin{bmatrix}
\toep(\bu) & \bX \\
\bX^* & \bW \end{bmatrix} \succeq \bf{0}, $$
with $\toep(\bu) = \bV\bD\bV^*$, $\bD=\diag(d_i)$, $d_i\ge 0$, and $\bV$ is a Vandermonde matrix. It follows that $\bX$ is in the range of $\bV$, hence $\bX=\bV\bB$ with the columns of $\bB^T$ given by $\bb_i$. Since $\bW$ is also PSD, $\bW$ can be written as $\bW = \bB^*\bE\bB$ where $\bE$ is also PSD. We now have
$$ \begin{bmatrix}
\toep(\bu) & \bX \\
\bX^* & \bW \end{bmatrix} =\begin{bmatrix}
\bV &   \\
  & \bB^* \end{bmatrix} \begin{bmatrix}
\bD & \bI \\
\bI & \bE \end{bmatrix} \begin{bmatrix}
\bV^* &  \\
 & \bB \end{bmatrix} \succeq \bf{0}, $$
which yields 
$ \begin{bmatrix}
\bD & \bI \\
\bI & \bE \end{bmatrix} \succeq \bf{0} $
and $\bE\succeq \bD^{-1}$ by the Schur complement lemma. Now observe
\begin{align*}
\trace(\bW)& = \trace(\bB^*\bE\bB)\geq  \trace(\bB^*\bD^{-1}\bB) \\
&= \trace(\bD^{-1}\bB\bB^*) = \sum_i d_i^{-1} \| \bb_i\|^2.
\end{align*}
Therefore,
\begin{align*}
\frac{1}{2}\trace(\toep(\bu))+\frac{1}{2} \trace(\bW) & =\frac{1}{2} \trace(\bD)+\frac{1}{2} \trace(\bW) \\
& \geq \sqrt{ \trace(\bD)\cdot \trace(\bW)} \\
& \geq \sqrt{\left(\sum_i d_i \right) \left( \sum_i d_i^{-1} \| \bb_i\|^2 \right) } \\
& \geq \sum \|\bb_i\| \geq \|\bX \|_\cA,
\end{align*}
which gives $\|\bX\|_\cT \geq \|\bX\|_\cA$. Therefore, $\|\bX\|_\cT = \|\bX\|_\cA$.
\end{proof}

\subsection{Proof of Proposition~\ref{dual_certificate}}\label{proof_dual_certificate}
\begin{proof} First, any $\bY$ satisfying \eqref{conditions} is dual feasible. We have
\begin{align*}
\|\bX^{\star}\|_{\cA} & \geq\|\bX^{\star}\|_{\cA}\|\bY\|_{\cA}^* \\
&\geq \langle \bY, \bX^{\star} \rangle_{\mathbb{R}}  =\langle \bY,  \sum_{k=1}^r c_k \ba(f_k)\bb_k^* \rangle_{\mathbb{R}} \\
& = \sum_{k=1}^r \mbox{Re}\left( c_k \langle \bY, \ba(f_k)\bb_k^* \rangle \right) \\
& = \sum_{k=1}^r \mbox{Re} \left( c_k \langle \bb_k, \bQ(f_k) \rangle \right) \\
& =  \sum_{k=1}^r \mbox{Re} \left( c_k \langle \bb_k, \bb_k \rangle \right)  = \sum_{k=1}^r c_k \geq \|\bX^{\star}\|_{\cA} .
\end{align*}
Hence $\langle \bY, \bX^{\star} \rangle_{\mathbb{R}} =\|\bX^{\star}\|_{\cA}$. By strong duality we have $\bX^{\star}$ is primal optimal and $\bY$ is dual optimal.

For uniqueness, suppose $\hat{\bX}$ is another optimal solution which has support outside $\mathcal{F}$. It is trivial to justify if $\hat{\bX}$ and $\bX^{\star}$ have the same support, they must coincide since the set of atoms with frequencies in $\mathcal{F}$ is independent. Let $\hat{\bX}=\sum_k \hat{c}_k \ba(\hat{f}_k)\hat{\bb}_k^*$. We then have the dual certificate
\begin{align*}
\langle\bY, \hat{\bX}  \rangle_{\mathbb{R}}  & = \sum_{\hat{f}_k\in\mathcal{F}}\mbox{Re}  \left( \hat{c}_k \langle \hat{\bb}_k, \bQ(\hat{f}_k) \rangle \right) +  \sum_{\hat{f}_l\notin\mathcal{F}}\mbox{Re}  \left( \hat{c}_l \langle \hat{\bb}_l, \bQ(\hat{f}_l) \rangle \right) \\
& <  \sum_{\hat{f}_k\in\mathcal{F}} \hat{c}_k + \sum_{\hat{f}_l \notin\mathcal{F}} \hat{c}_l  = \|\hat{\bX}\|_{\cA},
\end{align*}
which contradicts strong duality. Therefore the optimal solution of \eqref{primal} is unique.
\end{proof}

\subsection{Proof of Theorem~\ref{thm_mmv_rate}} \label{proof_mmv_rate}
 \begin{proof} We first record \cite[Proposition 1 and Theorem 1]{Bhaskar2013denoising} that applies to our atomic norm denoising formulation.
\begin{lemma}\label{convergence_rate}
If $\mathbb{E} \| \bN\|_{\cA}^{*}\leq \tau$, the solution to \eqref{equ:equivalent} satisfies that
\begin{align}\label{worst_case}
\mathbb{E} \left\Vert\hat{\boldsymbol{X}}-\boldsymbol{X}^{\star}\right\Vert_{\mathrm{F}}^{2}\leq 2\tau \left\Vert\boldsymbol{X}^{\star}\right\Vert_{\mathcal{A}}.
\end{align}
\end{lemma}

Lemma~\ref{convergence_rate} immediately implies that we can characterize the expected convergence rate of the atomic norm denoising algorithm \eqref{equ:equivalent} if the behavior of $\mathbb{E}\left\Vert\boldsymbol{N}\right\Vert_{\mathcal{A}}^{*}$ can be understood. According to the definition of dual norm, we can write $\left(\left\Vert\boldsymbol{N}\right\Vert_{\mathcal{A}}^{*}\right)^2$ as
\begin{align}
\left(\left\Vert\boldsymbol{N}\right\Vert_{\mathcal{A}}^{*}\right)^2
&=\sup_{f\in [0,1)}\left\Vert\boldsymbol{N}^{*}\boldsymbol{a}\left(f\right)\right\Vert_{2}^2 \nonumber \\
&=\sup_{f\in [0,1)} \sum_{l=1}^{L}\left\vert\frac{1}{\sqrt{n}}\sum_{i=1}^{n}\phi_{i,l}^{*}e^{j2\pi (i-1)f}\right\vert^{2} \nonumber \\
&=\sup_{f\in [0,1)} \sum_{l=1}^{L}\left\vert w_{l}\left(f\right)\right\vert^{2},\label{noise_dual_norm}
\end{align}
where $\phi_{i,l}$ is the $(i,l)$th entry of $\boldsymbol{N}$, and $w_{l}\left(f\right)\triangleq\frac{1}{\sqrt{n}}\sum_{i=1}^{n}\phi_{i,l}^{*}e^{j2\pi (i-1)f}$. For $f_1, f_2\in[0,1)$, by Bernstein's theorem \cite{schaeffer1941inequalities} and also a partial result in \cite[Appendix C]{Bhaskar2013denoising}, we can obtain that
\begin{equation}\label{bernstein}
 \left\vert w_{l}\left(f_{1}\right)\right\vert-\left\vert w_{l}\left(f_{2}\right)\right\vert \leq 4\pi n \left\vert f_{1}-f_{2}\right\vert\sup_{f\in [0,1)}\left\vert w_{l}\left(f\right)\right\vert.
\end{equation}
Therefore, we can write
\begin{align}
&\quad \sum_{l=1}^{L}\left\vert w_{l}\left(f_{1}\right)\right\vert^{2}-\sum_{l=1}^{L}\left\vert w_{l}\left(f_{2}\right)\right\vert^{2} \nonumber \\ 
&\le\sum_{l=1}^{L}\left(\left\vert w_{l}\left(f_{1}\right)\right\vert+\left\vert w_{l}\left(f_{2}\right)\right\vert\right)\left(4\pi n \left\vert f_{1}-f_{2}\right\vert\sup_{f\in [0,1)}\left\vert w_{l}\left(f\right)\right\vert\right)\label{eq38} \\
&\le 8\pi n \left\vert f_{1}-f_{2}\right\vert L \sup_{f\in [0,1)}\sum_{l=1}^{L}\left\vert w_{l}\left(f\right)\right\vert^{2}  \label{eq40}\\
&=8\pi n \left\vert f_{1}-f_{2}\right\vert L \left(\left\Vert\boldsymbol{N}\right\Vert_{\mathcal{A}}^{*}\right)^{2}, \nonumber
\end{align}
where \eqref{eq38} follows by plugging in \eqref{bernstein}, \eqref{eq40} follows from $\sum_{l=1}^{L}\sup_{f\in [0,1)}\left\vert w_{l}\left(f\right)\right\vert^{2} \leq L \sup_{f\in[0,1)}\sum_{l=1}^{L}\left\vert w_{l}\left(f\right)\right\vert^{2}  $, and the last equality follows from \eqref{noise_dual_norm}.

Let $D$ be a grid size that we will specify later, then by allowing $f_{2}$ to take any of the $D$ values on the grid points $\{0,\frac{1}{D},\dots,\frac{D-1}{D}\}$, we have
$$
\left(\left\Vert\boldsymbol{N}\right\Vert_{\mathcal{A}}^{*}\right)^{2}\le\max_{d=0,\ldots,D-1}\sum_{l=1}^{L}\left\vert w_{l}\left(\frac{d}{D}\right)\right\vert^{2}+\frac{4\pi nL}{D}\left(\left\Vert\boldsymbol{N}\right\Vert_{\mathcal{A}}^{*}\right)^{2}.
$$
Thus, $ \left\Vert\boldsymbol{N}\right\Vert_{\mathcal{A}}^{*} $ can be bounded as
\begin{equation}\label{equ:dualnormbound}
 \left\Vert\boldsymbol{N}\right\Vert_{\mathcal{A}}^{*} 
\leq \left(1-\frac{4\pi nL}{D}\right)^{-\frac{1}{2}}\left(\max_{d=0,\ldots,D-1}\sum_{l=1}^{L}\left\vert w_{l}\left(\frac{d}{D}\right)\right\vert^{2}\right)^{\frac{1}{2}}.
\end{equation}


Denote $Q_{d}\triangleq\frac{2}{\sigma^{2}}\sum_{l=1}^{L}\left\vert w_{l}\left(\frac{d}{D}\right)\right\vert^{2}$ which is a chi-squared random variable with $2L$ degrees of freedom. We first analyze $\mathbb{E}[ \left\Vert\boldsymbol{N}\right\Vert_{\mathcal{A}}^{*} ]$. 
From \eqref{equ:dualnormbound}, we have that
\begin{align}
&\mathbb{E}\left[\left\Vert\boldsymbol{N}\right\Vert_{\mathcal{A}}^{*}\right]\nonumber\\
&\leq \left(1-\frac{4\pi nL}{D}\right)^{-\frac{1}{2}}\left( \frac{\sigma^2}{2}\right)^\frac{1}{2} \left(\mathbb{E}\left[\max_{d=0,\ldots,D-1} Q_d \right]\right)^{\frac{1}{2}}\nonumber\\
&\le \left( \frac{\sigma^2}{2}\right)^\frac{1}{2}\left(1+\frac{8\pi nL}{D}\right)^{\frac{1}{2}}  \left(\mathbb{E}\left[\max_{d=0,\ldots,D-1} Q_d \right]\right)^{\frac{1}{2}}.\label{bound_dual}
\end{align}
Note that
\begin{align}
\mathbb{E}\left[\max_{d=0,\ldots,D-1}Q_{d}\right]&=\int_{0}^{\infty}P\left[\max_{d=0,\ldots,D-1}Q_{d}\ge t\right]dt \nonumber \\
&\le\delta+\int_{\delta}^{\infty}P\left[\max_{d=0,\ldots,D-1}Q_{d}\ge t\right]dt \nonumber\\
&\le\delta+D\int_{\delta}^{\infty}P\left[Q_{d}\ge t\right]dt, \label{bound_maxQ}
\end{align}
where the last line follows by the union bound. Recall the following lemma which bounds the tail behavior of a chi-squared random variable.
\begin{lemma}
Let $U$ be a standard chi-squared distribution of degrees of freedom $2L$, for any $\gamma>0$, we have
\begin{equation}\label{tail_chi_squared}
P \left[U\geq (1+\gamma+\frac{1}{2}\gamma^2)2L\right] \leq \exp\left( -\frac{L}{2}\gamma^2 \right).
\end{equation}
\end{lemma}

Plugging \eqref{tail_chi_squared} into \eqref{bound_maxQ}, we obtain
\begin{align}
&\mathbb{E}\left[\max_{d=0,\ldots,D-1}Q_{d}\right]\nonumber\\
&\leq \delta+2LD\int_{\delta}^{\infty}P\left[Q_{d}\ge (1+\gamma+\frac{1}{2}\gamma^2)2L\right]\left(1+\gamma\right)d\gamma \label{t_gamma}\\
&\le \delta+2LD\int_{-1+\frac{1}{L}\sqrt{L\left(\delta-L\right)}}^{\infty} \exp\left( -\frac{L}{2}\gamma^2 \right)\left(1+\gamma\right)d\gamma \label{use_lamma_1}\\
&=\delta+2\sqrt{L}D\sqrt{2\pi}\cdot Q\left(-\sqrt{L}+\sqrt{\left(\delta-L\right)}\right)\nonumber\\
&\quad +2De^{-\frac{L}{2}\left(-1+\frac{1}{L}\sqrt{L\left(\delta-L\right)}\right)^{2}}\label{use_Q_function}\\
&\le \delta+2\sqrt{L}D\sqrt{2\pi}\cdot \frac{1}{2}e^{-\frac{1}{2}\left(-\sqrt{L}+\sqrt{\left(\delta-L\right)}\right)^{2}}\nonumber\\
&\quad +2De^{-\frac{L}{2}\left(-1+\frac{1}{L}\sqrt{L\left(\delta-L\right)}\right)^{2}}\label{Q_relax}\\
&=\delta+\left(\sqrt{2\pi}\sqrt{L}+2\right)De^{-\frac{1}{2}\left(-\sqrt{L}+\sqrt{\left(\delta-L\right)}\right)^{2}}\nonumber,
\end{align} 
where \eqref{t_gamma} follows by letting $t= (1+\gamma+\frac{1}{2}\gamma^2)2L$, \eqref{use_lamma_1} follows from \eqref{tail_chi_squared}, \eqref{use_Q_function} follows by straight calculations using the definition of the $Q$ function,  and \eqref{Q_relax} follows by the Chernoff bound $Q\left(x\right)\le\frac{1}{2}e^{-\frac{x^{2}}{2}}$.

Let $-\frac{1}{2}\left(-\sqrt{L}+\sqrt{\left(\delta-L\right)}\right)^{2}=-\log{D}$, i.e. $\delta=2L+2\log{D}+2\sqrt{2L\log{D}}$, then we can obtain that
\begin{align*}
\mathbb{E}\left[\max_{d=0,\ldots,D-1}Q_{d}\right]\le 2L+2\log{D}+2\sqrt{2L\log{D}}+\sqrt{2\pi L}+2.
\end{align*}
Therefore, by plugging this into \eqref{bound_dual} and letting $D=8\pi nL\log{n}$, we obtain
\begin{align*}
&\mathbb{E}\left[\left\Vert\boldsymbol{N}\right\Vert_{\mathcal{A}}^{*}\right]\le \sigma  \left(1+\frac{1}{\log{n}}\right)^{\frac{1}{2}}\\
&\cdot  \left(L+\log{\left(\alpha L\right)}+\sqrt{2L\log{\left(\alpha L\right)}}+\sqrt{\frac{\pi L}{2}}+1\right)^{\frac{1}{2}},\\
\end{align*}
where $\alpha=8\pi n \log{n}$. The proof is completed by setting the right hand side as $\tau$.
 
\end{proof}

\subsection{Proof of Theorem~\ref{thm_mmv_cov}}\label{proof_mmv_cov}

\begin{proof}
As the trace norm is equivalent to the nuclear norm $\| \cdot\|_*$ for PSD matrices, we consider the equivalent algorithm 
\begin{equation}
\hat{\bu} = \argmin_{\bu}   \frac{1}{2}\left\Vert \mathcal{P}_{\Omega} \left( \mathcal{T}\left(\boldsymbol{u}\right) \right)-\boldsymbol{\Sigma}_{\Omega,L}\right\Vert_{F}^{2}  + \lambda \| \toep(\bu)\|_*.
\end{equation}

Denote the tangent space of $\toep(\bu)$ spanned its column and row space as $T$, and its orthogonal tangent space as $T^{\perp}$. Decompose the error term $\toep(\hat{\bu} - \bu^\star) = \boldsymbol{H}_1 + \boldsymbol{H}_2$ into two terms satisfying $\mbox{rank}(\bH_1)\leq 2r$ and $\bH_2\in T^{\perp}$ \cite{negahban2012unified}. Rephrasing straightforwardly~\cite[Lemma 1]{negahban2012unified}, we have that as long as $\lambda \geq \| \bSigma_{\Omega}^{\star}-\boldsymbol{\Sigma}_{\Omega , T} \|$, where $\bSigma^{\star}_{\Omega}= \mathcal{P}_{\Omega} \left( \mathcal{T}\left(\boldsymbol{u}^\star \right) \right)$,
\begin{equation}\label{tail}
\| \bH_2 \|_* \leq 3 \| \bH_1\|_*.
\end{equation} 

To obtain a reasonable regularization parameter $\lambda$, we utilize the following bound in \cite{bunea2012sample}.  
\begin{lemma}[\cite{bunea2012sample}]
Suppose that $\bx_l$ is a Gaussian random vector with mean zero and covariance $\bSigma$. Define the sample covariance matrix $\bSigma_L = \frac{1}{L}\sum_{l=1}^L \bx_l\bx_l^*$. Then with probability at least $1-L^{-1}$,
\begin{equation} \label{sample_cov}
\begin{split}
&\| \boldsymbol{\Sigma}_{L} - \boldsymbol{\Sigma} \| \\
&\leq C\max\left\{ \sqrt{\frac{r_{\text{eff}}(\bSigma) \log (Ln)}{L}}, \frac{r_{\text{eff}}(\bSigma) \log (Ln)}{L} \right\} \| \boldsymbol{\Sigma}\| \\
\end{split}
\end{equation}
for some constant $C$.
\end{lemma}

Instantiating \eqref{sample_cov} we have with probability at least $1-L^{-1}$,
\begin{align}\label{set_reg}
 &\| \bSigma_{\Omega}^{\star}-\boldsymbol{\Sigma}_{\Omega,L} \| \nonumber\\
 & \leq C\max\left\{ \sqrt{\frac{r_{\text{eff}}(\bSigma_{\Omega}^{\star}) \log (Ln)}{L}}, \frac{r_{\text{eff}}(\bSigma_{\Omega}^{\star}) \log (Ln)}{L} \right\} \| \bSigma_{\Omega}^{\star} \| \nonumber\\
 &: =\lambda.
\end{align}


%

Now by the triangle inequality,
\begin{align*}
\| \toep(\bu^\star)\|_* & = \| \toep(\bu^\star) - \toep(\hat{\bu})+ \toep(\hat{\bu}) \|_* \\
&\leq \| \bH \|_* + \|\toep(\hat{\bu}) \|_*,
 \end{align*}
 and by the optimality of $\hat{\bu}$:
 \begin{align*}
&\frac{1}{2} \| \mathcal{P}_{\Omega}(\toep(\hat{\bu})) - \bSigma_{\Omega,L}  \|_F^2 +\lambda \|\toep(\hat{\bu}) \|_* \\
&\leq  \frac{1}{2} \| \mathcal{P}_{\Omega}(\toep(\bu^\star)) - \bSigma_{\Omega,L}  \|_F^2 +\lambda \|\toep(\bu^\star) \|_* ,
 \end{align*}
 which gives
\begin{equation}\label{optimality_cond} 
\begin{split}
&\lambda \| \toep(\bu^\star) - \toep(\hat{\bu})\|_*\\
& \geq \frac{1}{2} \| \mathcal{P}_{\Omega}(\toep(\hat{\bu})) - \bSigma_{\Omega,L}  \|_F^2  - \frac{1}{2} \| \mathcal{P}_{\Omega}(\toep(\bu^\star)) - \bSigma_{\Omega,L}  \|_F^2.\\
\end{split}
\end{equation}
Further since
\begin{align*} 
&\quad\; \| \bSigma_{\Omega,L} - \mathcal{P}_{\Omega} (\toep(\hat{\bu}) ) \|_F^2  \\
& =  \| \bSigma_{\Omega,L} - \mathcal{P}_{\Omega} ( \toep(\bu^\star)) +\mathcal{P}_{\Omega} ( \toep(\bu^\star)) -  \mathcal{P}_{\Omega} ( \toep(\hat{\bu}))  \|_F^2 \\
& =  \| \bSigma_{\Omega,L} - \mathcal{P}_{\Omega} ( \toep(\bu^\star))   \|_F^2 + \|\mathcal{P}_{\Omega} ( \toep(\hat{\bu} - \bu^\star) )  \|_F^2 \\
&\quad + 2\langle \bSigma_{\Omega,L}  - \bSigma^{\star}_\Omega , \toep(\bu^\star) - \toep(\hat{\bu})  \rangle,
\end{align*}
which gives
\begin{align*}
&\|\mathcal{P}_{\Omega} ( \toep(\hat{\bu} - \bu^\star) )  \|_F^2 \\
& = \| \bSigma_{\Omega,L} - \mathcal{P}_{\Omega} (\toep(\hat{\bu}) ) \|_F^2 - \| \bSigma_{\Omega,L} - \mathcal{P}_{\Omega} ( \toep(\bu^\star))   \|_F^2 \\
&\quad - 2\langle \bSigma_{\Omega,L}  - \bSigma^{\star}_\Omega , \toep(\bu^\star) - \toep(\hat{\bu})  \rangle \\
& \leq 2\lambda  \| \toep(\bu^\star) - \toep(\hat{\bu})\|_* + 2\| \bSigma_{\Omega, L} -  \bSigma^{\star}_\Omega \| \cdot \|\toep(\bu^\star - \hat{\bu})\|_* \\
& \leq  4\lambda  \|\toep(\bu^\star - \hat{\bu})\|_* \\
& \leq 16 \lambda  \|\bH_1 \|_*    \leq 16 \lambda \sqrt{r} \| \bH_1 \|_{F} \leq 16 \lambda \sqrt{r} \| \toep(\hat{\bu} - \bu^\star)   \|_F,
\end{align*}
where the first inequality follows from \eqref{optimality_cond} and the Cauchy-Schwartz inequality, the second inequality follows from \eqref{set_reg}, and the third inequality follows from \eqref{tail}. We consider two cases:
\begin{itemize}
\item With full observation $\mathcal{P}_{\Omega} ( \toep(\hat{\bu} - \bu^\star) ) =\toep(\hat{\bu} - \bu^\star) $, we have
$  \| \toep(\hat{\bu} - \bu^\star)   \|_F \leq 16\lambda \sqrt{r}   . $
\item When $\Omega$ is a complete sparse ruler, we have
$$ \|\mathcal{P}_{\Omega} ( \toep(\hat{\bu} - \bu^\star) )  \|_F^2 \geq \| \hat{\bu} - \bu^\star \|_F^2 , $$ 
which gives
\begin{align*}
\| \hat{\bu} - \bu^\star \|_F^2 & \leq 16 \lambda \sqrt{r} \| \toep(\hat{\bu} - \bu^\star)   \|_F\\
& \leq 16\lambda \sqrt{rn} \|  \hat{\bu} - \bu^\star   \|_F.
\end{align*}
 Therefore we have $\frac{1}{\sqrt{n} }\| \hat{\bu} - \bu^\star \|_F \leq 16\lambda \sqrt{r} $.
\end{itemize}
\end{proof}

\subsection{ADMM Implementation of \eqref{equ:equivalent}}\label{ADMM}
In order to apply ADMM \cite{boyd2011distributed}, we reformulate \eqref{equ:equivalent} as
\begin{align*}
\min_{\boldsymbol{X}}&\; \frac{1}{2}\left\Vert\boldsymbol{X}-\boldsymbol{Z}\right\Vert^{2}_{F} + \frac{\tau}{2}\left(\mbox{Tr}\left(\mathcal{T}\left(\boldsymbol{u}\right)\right)+\mbox{Tr}\left(\boldsymbol{W}\right)\right)\\
 \mbox{s.t.}\quad &\; \boldsymbol{Y}=\begin{bmatrix}
\mathcal{T}\left(\boldsymbol{u}\right) & \boldsymbol{X} \\
\boldsymbol{X}^{*} & \boldsymbol{W} \end{bmatrix},\boldsymbol{Y}\succeq \mathbf{0} , 
\end{align*}
whose augmented Lagrangian can then be cast as
\begin{align*}
\Psi\left(\boldsymbol{X},\boldsymbol{u},\boldsymbol{W},\boldsymbol{\Lambda},\boldsymbol{Y}\right)=&\frac{1}{2}\left\Vert\boldsymbol{X}-\boldsymbol{Z}\right\Vert^{2}_{F} + \frac{\tau}{2}\left(\mbox{Tr}\left(\mathcal{T}\left(\boldsymbol{u}\right)\right)+\mbox{Tr}\left(\boldsymbol{W}\right)\right)\\
&+\left\langle \boldsymbol{\Lambda},\boldsymbol{Y}-\begin{bmatrix}
\mathcal{T}\left(\boldsymbol{u}\right) & \boldsymbol{X} \\
\boldsymbol{X}^{*} & \boldsymbol{W} \end{bmatrix} \right\rangle \\
&+\frac{\rho}{2}\left\Vert\boldsymbol{Y}-\begin{bmatrix}
\mathcal{T}\left(\boldsymbol{u}\right) & \boldsymbol{X} \\
\boldsymbol{X}^{*} & \boldsymbol{W} \end{bmatrix}\right\Vert^{2}_{F},
\end{align*}
where $\boldsymbol{Y}$, $\boldsymbol{W}$ and $\boldsymbol{\Lambda}$ are all Hermitian matrices. For notation simplicity, let $\boldsymbol{\Lambda}=\begin{bmatrix}
\boldsymbol{\Lambda}_{n\times n} & \boldsymbol{\Lambda}_{n\times L} \\
\boldsymbol{\Lambda}_{L\times n} & \boldsymbol{\Lambda}_{L\times L} \end{bmatrix}$, $\boldsymbol{Y}=\begin{bmatrix}
\boldsymbol{Y}_{n\times n} & \boldsymbol{Y}_{n\times L} \\
\boldsymbol{Y}_{L\times n} & \boldsymbol{Y}_{L\times L} \end{bmatrix}$. Then the update steps of ADMM are as follows
\begin{align*}
&\left(\boldsymbol{X}^{t+1},\boldsymbol{u}^{t+1},\boldsymbol{W}^{t+1}\right)=\argmin_{\boldsymbol{X},\boldsymbol{u},\boldsymbol{W}} \Psi\left(\boldsymbol{X},\boldsymbol{u},\boldsymbol{W},\boldsymbol{\Lambda}^{t},\boldsymbol{Y}^{t}\right);\\
&\boldsymbol{Y}^{t+1}=\argmin_{\boldsymbol{Y}\succeq\mathbf{0}}\Psi\left(\boldsymbol{X}^{t+1},\boldsymbol{u}^{t+1},\boldsymbol{W}^{t+1},\boldsymbol{\Lambda}^{t},\boldsymbol{Y}\right);\\
&\boldsymbol{\Lambda}^{t+1}=\boldsymbol{\Lambda}^{t}+\rho\left(\boldsymbol{Y}^{t+1}-\begin{bmatrix}
\mathcal{T}\left(\boldsymbol{u}^{t+1}\right) & \boldsymbol{X}^{t+1} \\
(\boldsymbol{X}^{t+1})^* & \boldsymbol{W}^{t+1} \end{bmatrix}\right),
\end{align*}
where the superscript $t$ denotes the $t$th iteration. Fortunately, closed-form solutions to the above updates exist and can be given as
\begin{align*}
&\boldsymbol{W}^{t+1}=\frac{1}{2}\boldsymbol{Y}_{L\times L}^{t}+\frac{1}{2}(\boldsymbol{Y}^{t}_{L\times L})^*+\frac{1}{\rho}\left(\boldsymbol{\Lambda}_{L\times L}^{t}-\frac{\tau}{2}\boldsymbol{I}\right);\\
&\boldsymbol{X}^{t+1}=\frac{1}{2\rho+1}\left(\boldsymbol{Z}+2(\boldsymbol{\Lambda}^{t}_{L\times n})^*+\rho\boldsymbol{Y}_{n\times L}^{t}+\rho(\boldsymbol{Y}^{t}_{L\times n})^*\right);\\
&\boldsymbol{u}^{t+1}=\frac{1}{\rho}\cdot\boldsymbol{\Upsilon} \cdot\mbox{conj}\left(\mathcal{G}\left(\boldsymbol{\Lambda}_{n\times n}^{t}\right)+\rho\mathcal{G}\left(\boldsymbol{Y}_{n\times n}^{t}\right)-\frac{\tau}{2}n\boldsymbol{e}_{1}\right),
\end{align*}
where $\mbox{conj}\left(\cdot\right)$ means the conjugate operation on each entry of a vector or a matrix, $\boldsymbol{e}_{1}$ is the first vector in the standard basis, $\boldsymbol{a}=\mathcal{G}\left(\boldsymbol{A}\right)$ is a mapping from a matrix to a vector where the $i$th entry in $\boldsymbol{a}$ is the sum of all the entries ${A}_{p,q}$'s of $\boldsymbol{A}$ satisfying $q-p+1=i$, and $\boldsymbol{\Upsilon}$ is a diagonal matrix with diagonal entries
$\Upsilon_{i,i}=\frac{1}{n-i+1}$, $i=1,\ldots,n$.

Let $\boldsymbol{\Xi}^{t}=\begin{bmatrix}
\mathcal{T}\left(\boldsymbol{u}^{t+1}\right) & \boldsymbol{X}^{t+1} \\
\boldsymbol{X}^{*t+1} & \boldsymbol{W}^{t+1} \end{bmatrix}-\frac{1}{\rho}\boldsymbol{\Lambda}^{t} = \boldsymbol{U}^{t} \mbox{diag}(\{\sigma_i^t\})(\boldsymbol{U}^{t})^*$ be its eigenvalue decomposition, then the update of $\boldsymbol{Y}$ can be given as 
$$\boldsymbol{Y}^{t+1}=\boldsymbol{U}^{t}\mbox{diag}(\{\sigma_i^t\}_{+})(\boldsymbol{U}^{t})^*.$$ 


We run the above iterations until both primal and dual residuals satisfy the pre-set tolerance level.

\bibliographystyle{IEEEtran} 
\bibliography{bibfileSparseMatrixPencil_atomic}

\end{document}